\documentclass{article}

\usepackage{amsfonts} %%------ AMS Fonts package ------------------
\usepackage{amsmath} %%------- Graphics package -------------------
\usepackage{graphicx}   %% --- Graphics package -------------------
\usepackage{color}  %%-------- Color package ----------------------
\usepackage{wrapfig} %%------- Wrapping text around a figure ------
\usepackage{epsfig} %%-------- EPS support? -----------------------
\usepackage{wasysym} %%------- ??? --------------------------------
\usepackage{pifont} %% ------- ??? --------------------------------
\usepackage{mathrsfs} %%------ Special math fonts -----------------
\usepackage{array} %%--------- Table formatting aids---------------
\newcommand{\TT}{\scriptscriptstyle} %%----------- Smallest size---
\newcommand{\vb}[1]{{\boldsymbol {#1}}} %%-------- For vector face-

\definecolor{brown}{rgb}{0.65,0.325,0} %%--------- Brown color ----
\definecolor{purple}{rgb}{1,0.5,0.75} %%---------- Blue color -----
\begin{document}

\title{Green Functions For Wave Propagation on a 5D manifold and the
       Associated Gauge Fields Generated by a Uniformly Moving
       Point Source}

\author{ I. Aharonovich$^{a}$
         and
         L. P. Horwitz$^{abc}$ \\
         \\
         $^{a}$ Bar-Ilan University, Department of Physics, Ramat Gan, Israel.  \\
         $^{b}$ Tel-Aviv University, School of Physics, Ramat Aviv, Israel. \\
         $^{c}$ College of Judea and Samaria, Ariel, Israel. \\
       }

%------------------------------------------------------------------
\maketitle
\rightline{TAUP 2282/06}
%------------------------------------------------------------------

\begin{abstract}
Gauge fields associated with the manifestly covariant dynamics of
particles in $(3,1)$ spacetime are five-dimensional. We provide
solutions of the classical 5D gauge field equations in both $(4,1)$
and $(3,2)$ flat spacetime metrics for the simple example of a
uniformly moving point source. Green functions for the 5D field
equations are obtained, which are consistent with the solutions for
uniform motion obtained directly from the field equations with free
asymptotic conditions.
\end{abstract}


%%\tableofcontents

\section{Introduction}
%%%    \input{UMP_Introduction}
% =============================================================== %
% --------------------------------------------------------------- %
% ----------------------> Introduction <------------------------- %
% --------------------------------------------------------------- %
% =============================================================== %
Maxwell electrodynamics arises in a natural way in the study of
quantum dynamical evolution of particles in 3D space. The
non-relativistic Schr\"odinger equation has a form which invariant
under the transformation
\begin{align}
    \Psi_{t}(x) & \rightarrow e^{i \Lambda(x,t)} \Psi_{t}(x)
\end{align}
when the so-called gauge compensation fields are added to the space
and time derivation. One finds in this method how the 3D dynamics
associated with non-relativistic theory results in a 4D gauge field,
which has an $O(3,1)$ invariance for the \emph{homogeneous} field
equations.

In a similar way, the manifestly covariant Stueckelberg
Schr\"odinger \cite{Stueckelberg1941} equation (discussed below in
Section \ref{sec:fundamentals}) induces five gauge fields. Here, we
study mathematical and physical properties of these 5D gauge fields.
The work is divided as follows.

%%
%%
%% Section #1 review
%%
%%
In Section \ref{sec:fundamentals}, an overview of 5D off-mass-shell
gauge field theory based on the framework of Stueckelberg
\cite{Stueckelberg1941,HorPir1973,SaadHorArsh1989} is given. 5D
gauge theories also arise in other studies, such as a special case
of higher dimensional relativistic dynamics and electrodynamics (cf.
\cite{Kazinski2002,GalTsov2002}), or modern Kaluza-Klein type
theories (cf. \cite{Wesson1999} and references therein). In this
paper, we concentrate on the construction which from the
Stueckelberg framework (cf. \cite{HorPir1973,SaadHorArsh1989}).

Previous studies of the fields have been conducted (cf.
\cite{OronHorwitz2001,HorKatzOron2004,MCLand1997}) using certain
types of Green functions (GF's). Since the field equation in higher
dimension admits many types of GF's, in order to gain some insight
into criteria for selecting useful ones, we study here a direct
solution by Fourier transform, for the special case of a uniformly
moving point source (UMS).

%%
%%
%% Section #2 review
%%
%%
In Section \ref{sec:solutions}, a derivation of 5D gauge fields
generated by a uniformly moving point source is given, for both
$(4,1)$ and $(3,2)$ metrics, followed by classification to
regions of source motion, namely, spacelike and timelike. The wave
equations are solved with asymptotically free conditions, in which
the boundary value of the fields at infinity vanish pointwise.

%%
%%
%% Section #3 review
%%
%%
In Section \ref{sec:green_functions}, a derivation is given of the
principal part GF's consistent with the fields generated by a
uniformly moving source of Section \ref{sec:solutions}.

The GF's obtained agree with a particular form of fundamental
solutions of 5D wave equations, found in, for example ref.
\cite{Kythe1996,Gelfand1964_1}, i.e.,
\begin{align}
    g(x,\tau)
    & =
        \lim\limits_{\epsilon \to 0^{+}}
        \dfrac{1}{4\pi^2}
        G_{\epsilon}(x^2 + \sigma_5 \tau^2)
    \\
    \label{eq:G_epsilon_function}
    G_{\epsilon}(y)
    & =
        \dfrac{\partial}{\partial \epsilon}
        \dfrac{\theta(-\sigma_5 y + \epsilon)}
              {\sqrt{-\sigma_5 y + \epsilon}}
\end{align}
where $\sigma_5$ determines the metric
signature, $\pm 1$ for $(4,1)$ and $(3,2)$ metrics, respectively.

Our present study differs from the previous literature in the
following:
\begin{itemize}
    \item{The GF's carry the group symmetry in all coordinates,
          whereas normally, only the $t$ retarded solutions are considered.}

    \item{The GF's are treated in a unified manner in both
            $(4,1)$ and $(3,2)$ metrics.
         }

    \item{We shall show that the derivative present in
          \eqref{eq:G_epsilon_function} is useful in
          regularizing the fields, whereas non derivative
          forms (cf. \cite{Kazinski2002} and references
          therein) have an additional infinite part, which
          may be removed by other methods such as
          Hadamard's \emph{finite part} (cf.
          \cite{Zemanian1965})}

    \item{We study, in particular, the properties of the gauge
          fields, derived both from the GF's and a more direct
          method, generated by a uniformly moving point source.
    }
\end{itemize}

% =============================================================== %
% --------------------------------------------------------------- %
% ----------------------> Fundamentals <------------------------- %
% --------------------------------------------------------------- %
% =============================================================== %
\section{Fundamentals} \label{sec:fundamentals}

An offshell classical and quantum electrodynamics has been
constructed \cite{SaadHorArsh1989} from a fundamental theory of
relativistic dynamics of 4D particles, termed \emph{events}, in a
framework first derived by Stueckelberg
\cite{Stueckelberg1941,HorPir1973}.

Stueckelberg defined a Lorentz invariant Hamiltonian-like generator
of evolution, over 8D phase space, parameterized by a Lorentz
invariant $\tau$, in both classical and quantum relativistic
mechanics. Solutions of the relativistic quantum two body bound
state problem agree, up to relativistic corrections
\cite{ArshHor1989}, with solutions of the non-relativistic
Schr\"odinger equation. The experiments of Lindner, et. al.
\cite{Lindner2005}, moreover, showing quantum interference in time
can be explained in a simple and consistent way in the framework of
this theory \cite{Horwitz2005}, and provides strong evidence that
the time $t$ is a quantum observable, as required in this framework.

In the classical manifestly covariant theory, the Hamiltonian
of a free particle is given by
\begin{align}
    \label{eq:stueckelberg_classical}
    K = \dfrac{p_{\mu} p^{\mu}}{2M}
\end{align}
where $x^{\mu} = [ct, \vb{x}]$ and $p^{\mu} = [E/c, \vb{p}]$. A
simple model for an interacting system is provided by the potential
model
\begin{align}
    \label{eq:stueckelberg_classical_with_V}
    K & =
        \dfrac{p_{\mu} p^{\mu}}{2M}
        +
        V(x)
\end{align}
The equation are
\begin{align}
    \label{eq:hamilton_equations_classical_K}
    \dot{x}^{\mu} =  \dfrac{\partial K}{\partial p_{\mu}} =  \dfrac{1}{M} p^{\mu} \qquad \qquad
    \dot{p}^{\mu} = -\dfrac{\partial K}{\partial x_{\mu}} = -\dfrac{\partial V}{\partial x^{\mu}}
\end{align}
It follows from \eqref{eq:hamilton_equations_classical_K} that
\begin{align*}
    \vb{v}
    & =
        \dfrac{d\vb{x}}{dt}
    =
        \dfrac{\dot{\vb{x}}}{\dot{t}}
    =
        \dfrac{\vb{p}}{E}
\end{align*}
which is the standard formula obtained for velocity in special
relativity (we take $c=1$ in the following).

Horwitz and Piron \cite{HorPir1973} generalized the framework
to many-body systems, and gave $\tau$ the physical meaning of a
\emph{universal historical time}, correlating events in spacetime.

The general many-body, $\tau$ invariant, classical evolution
function is defined as
\begin{align}
    \label{eq:Stueckelberg_Classical_Hamiltonian_Many_Particles}
    K = \sum_{\TT n = 1}^{N}
        \dfrac{1}{2M_{n}} \eta_{\mu \nu} P_{n}^{\mu} P_{n}^{\nu} +
        V(x_{1}, x_{2}, ... , x_{N})
\end{align}
where $\eta_{\mu \nu} = diag(-,+,+,+)$ and $n$ sums over all
particles of the system, and, in this case, we have taken the
potential function $V$ not to be a function of momenta or $\tau$.
The classical equations of motion, for a single particle system in
an external potential $V(x)$, are similar to the non-relativistic
Hamilton equations, with, in addition, motion and "momentum" along
the $t$ axis:
\begin{align}
    \dot{x}_{n}^{\mu} =  \dfrac{\partial K}{\partial p_{n \; \mu}} =   \dfrac{1}{M_n} p_{n}^{\mu}                    \qquad \qquad
    \dot{p}_{n \; \mu} = -\dfrac{\partial K}{\partial x_{n}^{\mu}} = - \dfrac{\partial V}{\partial x_{n}^{\mu}}
\end{align}
In the usual formulation of relativistic dynamics (cf.
\cite{Rindler1991}), the energy-momentum is constrained to a
\emph{mass-shell} defined as
\begin{align}
    p^{\mu} p_{\mu} = \mathbf{p}^2 - E^2 = - m^2
\end{align}
where $m$ is a given fixed quantity, a property of the particle. In
the Stueckelberg formulation, however, the event mass is generally
unconstrained. Since in \eqref{eq:stueckelberg_classical_with_V},
the value of $K$ is absolutely conserved, $p_{\mu} p^{\mu} = -m^2$
is constant only in the special case where
\begin{align*}
    \dfrac{d}{d\tau} V(x) =
        \left[
            \dot{x}^{\mu}
            \dfrac{\partial}{\partial x^{\mu}}
            +
            \dfrac{\partial}{\partial \tau}
        \right]
        V(x) =
        0
\end{align*}
In this case, the particle remains in a specific mass shell, which
may or may not coincide with its so-called \emph{Galilean target
mass}, usually denoted by $M$\footnote{In the non-relativistic
limit, the mass distribution converges to a single point; one may
choose the parameter $M$ to have this \emph{Galilean target mass
value} \cite{Burakovsky1996}. We shall assume that $M$ has this
value in the following. }. In the general case, however, $p^{\mu}
p_{\mu} \equiv -m^2$ is a dynamical (Lorentz invariant) property,
which may depend on $\tau$. The relation between $\tau$ and the
proper time $s$, in the model of eq.
\eqref{eq:Stueckelberg_Classical_Hamiltonian_Many_Particles}, is
given by
\begin{align}
    ds^2 & \equiv - dx^{\mu} dx_{\mu}     = - \dot{x}^{\mu} \dot{x}_{\mu} d\tau^2 =
                  - \dfrac{1}{M^2} p^{\mu} p_{\mu} d\tau^2 =
                  \dfrac{m^2}{M^2} d\tau^2
\end{align}
Thus, the proper time $ds$, and \emph{universal time} $d\tau$, are
related through the ratio between the dynamical Lorentz invariant
mass $m$, and the \emph{Galilean target mass} $M$. If $V(x)$ goes to
zero asymptotically, then it becomes constant. Since this asymptotic
value is usually what is measured in experiment, we may assume that
it takes on the value of the Galilean target mass. Although there
are no detailed models at present, one assumes that there is a
stabilizing mechanism (for example, self-interaction or, in terms of
statistical mechanics and condensation phenomenon
\cite{Burakovsky1996}) which brings the particle, at least to a good
approximation, to a defined mass value \cite{Burakovsky1996}, such
that
\begin{align*}
    K = \dfrac{1}{2M} p^{\mu} p_{\mu} = \dfrac{-m^2}{2M} = - \dfrac{M}{2}
\end{align*}

For the quantum case, for which $P^{\mu}$ is represented by $-i
\partial / \partial x_{\mu}$, the Stueckelber Schr\"odinger equation
is taken to be (we take $\hbar = 1$ in the following)
\begin{align}
    \label{eq:stueckelberg_hamiltonain_quantum}
    i \dfrac{\partial \Psi_{\tau}(x)}{\partial \tau} = K \Psi_{\tau}(x)
\end{align}
The Stueckelberg classical and quantum relativistic dynamics have
been studied for various systems in some detail, including the
classical relativistic Kepler problem \cite{HorPir1973} and the
quantum two body problem for central potential \cite{ArshHor1989}.

% ================================================================================================================= %
% ----------------------------------------------------------------------------------------------------------------- %
% ----------------------> OSE Electrodynamics <-------------------------------------------------------------------- %
% ----------------------------------------------------------------------------------------------------------------- %
% ================================================================================================================= %
\subsection{Off-Shell Electrodynamics}
Pre-Maxwell off-shell electrodynamics is constructed in a similar fashion to the
construction of Maxwell electrodynamics from the Schr\"odinger equation
\cite{SaadHorArsh1989}.

Under the local gauge transformation
\begin{align}
    \label{eq:gauge_transformation_definition}
    \Psi'_{\tau}(x) = e^{- i e_0 \chi(x,\tau)} \Psi_{\tau}(x)
\end{align}
5 compensation fields $a^{\alpha}(x,\tau)$ ($\alpha \in
\{0,1,2,3,5\}$) are implied, such that with the transformation
\begin{align*}
    a'_{\TT \alpha}(x,\tau) = a_{\alpha}(x,\tau) - \partial_{\alpha} \chi(x,\tau)
\end{align*}
the following modified Stueckelberg-Schr\"odinger equation remains form invariant
\begin{align}
    \label{eq:stueckelberg_schrodinger_after_gauge}
    \left[ i \dfrac{\partial}{\partial \tau} + e_0 a_{5}(x,\tau) \right] \Psi_{\tau}(x) =
    \dfrac{1}{2M}
    \left[
        (p^{\mu} - e_0 a^{\mu})
        (p_{\mu} - e_0 a_{\mu})
    \right]
    \Psi_{\tau}(x)
\end{align}
under the transformation \eqref{eq:gauge_transformation_definition}.

We can see this by observing the following relations:
\begin{align*}
    \left[ p_{\mu} - e_0 a'_{\mu} \right] \Psi' & =
        \left[- i \dfrac{\partial}{\partial x^{\mu}} - e_0 \left(a_{\mu} - \dfrac{\partial}{\partial x^{\mu}} \chi \right) \right]
        e^{-i e_0 \chi} \Psi = \\
    & =
        \left[
            - e_0 \dfrac{\partial}{\partial x^{\mu}} \chi  - i \dfrac{\partial \Psi}{x^{\mu}} -
            e_0 \left(a_{\mu} - \dfrac{\partial \chi}{\partial x^{\mu}} \right)
        \right]
        e^{-i e_0 \chi} \Psi  = \\
    & =
        e^{-i e_0 \chi}  [P_{\mu} - e_0 a_{\mu}] \Psi \\
        \\
    [i \dfrac{\partial}{\partial \tau} + e_0 (a_{5} - \dfrac{\partial \chi}{\partial \tau})]
    e^{-i e_0 \chi} \Psi  & =
        [
            e_0 \dfrac{\partial \chi}{\partial \tau}
            +   i \dfrac{\partial \Psi}{\partial \tau}
            +   e_0 ( a_{5} - \dfrac{\partial \chi}{\partial \tau} )
        ]
            e^{- i e_0 \chi} \Psi = \\
    & =
        e^{-i e_0 \chi}
        [
            i \dfrac{\partial}{\partial \tau}
            +   e_0 a_5
        ]   \Psi
\end{align*}
The result is then, of the same form as for the usual $U(1)$ gauge
compensation argument for the non-relativistic Schr\"{o}dinger
equation. Thus, the classical (and quantum) evolution function for a
particle, under an external field, assumed to be given by
\begin{align}
    K =
        \dfrac{1}{2M}
        \left[
            p - e_0 a(x,\tau)
        \right]^2
        -
        e_0 a^{5}(x,\tau)
\end{align}
(where we have used the shorthand notation of $x^2 = x_{\mu} x^{\mu}$)
and the corresponding Hamilton equations are
\begin{align}
    \label{eq:Hamilton_Equations_1}
    \dot{x}^{\mu}(\tau) & =
        \dfrac{\partial K}{\partial p_{\mu}} =
        \dfrac{1}{M}
        \left[
            p^{\mu} - e_0 a^{\mu}
        \right] \\
    \label{eq:Hamilton_Equations_2}
    \dot{p}^{\mu}(\tau) & =
        - \dfrac{\partial K}{\partial x_{\mu}} =
        \dfrac{e_0}{M}
        \left(
            p - e_0 a(x,\tau)
        \right)_{\nu}
        \partial^{\mu} a^{\nu}(x,\tau) +
        e_0 \partial^{\mu} a^{5}(x,\tau)
\end{align}
Here, $e_0$ is proportional to the Maxwell charge $e$ through a
dimensional constant, which is derived below. Second order equations
of motion for $x^{\mu}(\tau)$, a generalization of the usual Lorentz
force, follow from the Hamilton equations
\eqref{eq:Hamilton_Equations_1} and \eqref{eq:Hamilton_Equations_2}
\cite{SaadHorArsh1989}
\begin{align}
    \label{eq:5D_Lorentz_force}
    M \ddot{x}^{\mu} = e_0 \dot{x}^{\nu} f_{\; \nu}^{\mu} + e_0 f_{\; 5}^{\mu}
\end{align}
where for $\alpha,\beta = 0,1,2,3,5$ the antisymmetric tensor
\begin{align}
    \label{eq:ose_fields_from_potentials}
    f^{\alpha \beta} \equiv \partial^{\alpha} a^{\beta} - \partial^{\beta} a^{\alpha}
\end{align}
is the 5D field tensor. Moreover, second order wave equation for the
fields $f^{\alpha \beta}$ can be derived from a Lagrangian density
as follows \cite{SaadHorArsh1989}:
\begin{align}
    \label{eq:ose_field_langarangian}
    \mathscr{L} =  -\dfrac{\lambda}{4} f_{\alpha \beta} f^{\alpha \beta} - e_0 a_{\alpha} j^{\alpha}
\end{align}
which produces the wave equation
\begin{align}
    \label{eq:ose_fields_wave_equation}
    \lambda \partial_{\alpha} f^{\alpha \beta} = e_0 j^{\beta}
\end{align}
$\lambda$ is a dimensional constant, which will be shown below to have dimensions of length.
The sources $j^{\beta}(x,\tau)$ depend both on spacetime and on $\tau$, and obey
the continuity equation
\begin{align}
    \label{eq:continuity}
    \partial_{\alpha} j^{\alpha} = \partial_{\mu} j^{\mu} + \partial_{\tau} \rho = 0
\end{align}
where $j^{5} \equiv \rho$ is a Lorentz invariant \emph{spacetime
density of events}. This equation follows from
\eqref{eq:stueckelberg_schrodinger_after_gauge} for
\begin{align*}
    \rho_{\tau}(x) & = \Psi^{*}_{\tau}(x) \Psi_{\tau}(x)
    \\
    j^{\mu}_{\tau}(x) & =
        - \dfrac{i}{2M}
        \left[
            \Psi^{*}_{\tau}(x)
            \left(
                i \partial^{\mu} - e_{0} a^{\mu}(x,\tau)
            \right)
            \Psi_{\tau}(x)
            +
            c.c.
        \right]
\end{align*}
as we discuss below, and also the classical from the argument given
below.

\subsubsection{Currents of point events}
Maxwell current conservation, for point charges, can be derived (cf. \cite{Jackson1995})
by defining the current of a point charge as
\begin{align}
    \label{eq:maxwell_current_of_point_particle}
    J^{\mu}(x) = e \int_{-\infty}^{+\infty} ds \, \dot{z}^{\mu}(s) \delta^4 [x - z(s)]
\end{align}
In that case, $s$ is the proper time, and $z^{\mu}(s)$ the world-line of the point charge
(for free motion, $s$ may coincide with $\tau$), and $\dot{z}^{\mu}(s) = \dfrac{d}{ds} z^{\mu}(s)$.
Then,
\begin{align}
    \label{eq:maxwell_current_conservation_proof}
    \partial_{\mu} J^{\mu} =
        - e \int_{-\infty}^{+\infty} ds \, \dfrac{d}{ds} \delta^4 [x - z(s)] =
        - e \, \lim \limits_{L \rightarrow +\infty} \delta^4 [x - z(s)] \Bigg|_{-L}^{+L}
\end{align}
which vanishes if $z^{\mu}(s)$ (or, for example, just the time
component $z^{0}(s)$) becomes infinite for $s \rightarrow \pm
\infty$, and the observation point $x^{\mu}$ is restricted to a
bounded region of spacetime, e.g., the laboratory. We therefore,
with Jackson \cite{Jackson1995}, identify $J^{\mu}$ as the Maxwell current. We see that
this current is a functional on the world line, and the usual notion
of a "particle" corresponds to this functional on the world line.

If we identify $\delta^4 [x - z(s)]$ with a density $\rho_{s}(x)$
and the local (in $\tau$) current $\dot{z}^{\mu}(s) \delta^4 [x -
z(s)]$ with a local current $j^{\mu}_{s}(x)$
\begin{align}
    \label{eq:ose_current_of_point_event}
    \rho_{s}(x)      = \delta^4 [x - z(s)] \qquad \qquad
    j^{\mu}(x,s) = \dot{z}^{\mu}(s) \delta^4 [x - z(s)]
\end{align}
then the relation
\begin{align*}
    \dfrac{d}{ds} \delta^4 [x - z(s)] = -\dot{z}^{\mu}(s) \partial_{\mu} \delta^4 [x - z(s)]
\end{align*}
used in the above demonstration in fact corresponds to the
conservation law (reverting to the more general parameter $\tau$ in
place of the proper time $s$) \eqref{eq:continuity}
\begin{align}
    \label{eq:ose_current_conservation}
    \partial_{\mu} j^{\mu}(x,\tau) + \partial_{\tau} \rho(x,\tau) = 0
\end{align}
What we call the \emph{pre-Maxwell} current of a point \emph{event}
is then defined as
\begin{align}
    \label{eq:point_source_current}
    j^{\alpha}(x,\tau) = \dot{z}^{\alpha}(\tau) \delta^4 [x - z(\tau)]
\end{align}
where $j^{5}(x,\tau) \equiv \rho(x,\tau)$ and $\dot{z}^{5}(\tau)
\equiv 1$ (since $z^{5}(\tau) \equiv \tau$). Integrating
\eqref{eq:ose_fields_wave_equation} over $\tau$, we recover the
standard Maxwell equations for Maxwell fields defined by
\begin{align}
    A^{\mu}(x) & = \int a^{\mu}(x,\tau) d\tau
\end{align}
We therefore call the fields $a^{\mu}(x,\tau)$ \emph{pre-Maxwell
fields}. Thus, Maxwell theory is properly contained in the more
general pre-Maxwell theory.

For the quantum theory, a real positive definite density function $\rho_{\tau}(x)$
can be derived from the Stueckelberg-Schr\"odinger equation \eqref{eq:stueckelberg_hamiltonain_quantum}
\begin{align}
    \rho_{\tau}(x) = |\Psi_{\tau}|^2 = \Psi^{*}_{\tau} (x) \Psi_{\tau}(x)
\end{align}
which can be identified with the $\rho(x,\tau) = \delta^4[x - z(\tau)]$ in the classical (relativistic) limit.
The continuity equation \eqref{eq:ose_current_conservation} is then satisfied for the gauge invariant
currents
\begin{align}
    j^{\mu}_{\tau}(x) =
        - \dfrac{1}{2M}
        \left[
            \Psi^{\ast}_{\tau}(x) (i \partial^{\mu} - e_0 a^{\mu}(x,\tau)) \Psi_{\tau}(x)  +
            \text{c.c.}
        \right]
\end{align}

Combining \eqref{eq:maxwell_current_of_point_particle} with \eqref{eq:ose_current_of_point_event}, we obtain
\begin{align}
    \label{eq:ose_maxwell_current_integral}
    J^{\mu}(x) = e \int_{-\infty}^{+\infty} j^{\mu}(x,\tau) \, d\tau
\end{align}
This is a restatement of the 5D continuity equation
\eqref{eq:ose_current_conservation} and provides a connection
between pre-Maxwell \emph{OSE} and Maxwell electrodynamics. It also
follows from \eqref{eq:stueckelberg_hamiltonain_quantum} that
\begin{align}
    &
    \dfrac{\partial \rho}{\partial t} + \partial_{\mu} j^{\mu} = 0
\end{align}
where $\rho = |\Psi_{\tau}(x)|^2$, as for the classical case.
\\

\subsubsection{The wave equation}
From equations \eqref{eq:ose_fields_wave_equation} and \eqref{eq:ose_fields_from_potentials} one can derive
the wave equation for the potentials $a^{\TT\alpha}(x,\tau)$:
\begin{align}
    \label{eq:ose_potentials_wave_equation_before_gauge}
    \lambda \partial_{\beta} \partial^{\beta} a^{\alpha} -
    \lambda \partial^{\alpha} (\partial_{\beta} a^{\beta}) = e_0 \, j^{\alpha}
\end{align}
Under the generalized Lorentz gauge $\partial_{\beta} a^{\TT\beta} =
0$, the wave equation takes the simpler form
\begin{align}
    \label{eq:ose_potentials_wave_equation_after_gauge}
    \lambda \partial_{\beta} \partial^{\beta} a^{\alpha} =
    \lambda \left[ \Box^2 a^{\alpha} + \sigma_5 \dfrac{\partial^2 a^{\alpha}}{\partial \tau^2} \right] =
    e_0 \, j^{\alpha}(x,\tau)
\end{align}
where a $5^{th}$ diagonal metric component can take either signs
$\sigma_5 = \pm 1$, corresponding to $O(4,1)$ and $O(3,2)$
symmetries of the homogeneous field equations, respectively.

Integrating \eqref{eq:ose_potentials_wave_equation_after_gauge} with respect to $\tau$,
and assuming that $\lim\limits_{\tau \rightarrow \pm \infty} \partial_{\tau} a^{\alpha}(x,\tau) = 0$,
we obtain
\begin{align*}
    \lambda
    \int_{-\infty}^{+\infty} d\tau \,
    \left[ \Box^2 a^{\alpha} + \sigma_5 \dfrac{\partial^2 a^{\alpha}}{\partial \tau^2} \right] =
    \dfrac{e_0}{e} J^{\alpha}(x)
\end{align*}
Identifying
\begin{align}
    \label{eq:maxwell_from_pre_maxwell_potential}
    A^{\mu}(x) = \int_{-\infty}^{+\infty} d\tau \, a^{\mu}(x,\tau)
\end{align}
we obtain
\begin{align*}
    \lambda \Box^2 A^{\mu}(x) = \dfrac{e_0}{e} J^{\mu}(x)
\end{align*}
(where we've restricted our attention to $\mu = 0,1,2,3$), from which
a relation between $e_0$, $\lambda$ and the Maxwell charge $e$ can be obtained:
\begin{align}
    \label{eq:ose_relation_lambda_e0_and_e}
    e = \dfrac{e_0}{\lambda}
\end{align}

Therefore, the Maxwell electrodynamics is properly contained in the
5D electromagnetism.

\subsubsection{A note about units}
In natural units ($\hbar = c = 1$), the Maxwell potentials $A^{\mu}$ have units of $1/L$.
Therefore, the pre-Maxwell OSE potentials $a^{\alpha}$ have units of $1/L^2$, and
in order to maintain the action integral
\begin{align}
    S = \int_{-\infty}^{+\infty} \mathscr{L} \, d\tau \, d^4x
\end{align}
dimensionless, the coefficient $\lambda$ in \eqref{eq:ose_field_langarangian} must have units of $L$, forcing
$e_0$ to have units of $L$ as well.

The Fourier transform of the pre-Maxwell OSE fields
\begin{align}
    \tilde{a}^{\mu}(x,s) = \int_{-\infty}^{+\infty} e^{is\tau} a^{\mu}(x,\tau) \, d\tau
\end{align}
and equation \eqref{eq:maxwell_from_pre_maxwell_potential} suggest that the Maxwell potentials and fields
are the \emph{zero mode} of the pre-maxwell OSE fields, with respect to the $\tau$ axis, i.e.,
\begin{align}
    \label{eq:maxwell_field_from_zero_mode}
    A^{\mu}(x) = \tilde{a}^{\mu}(x,s) |_{s = 0}
\end{align}

% ================================================================================================================= %
% ----------------------------------------------------------------------------------------------------------------- %
% ----------------------> Solutions of the equations <------------------------------------------------------------- %
% ----------------------------------------------------------------------------------------------------------------- %
% ================================================================================================================= %
\subsection{Solutions of the wave equation}
The classical 5D wave-equation
\eqref{eq:ose_potentials_wave_equation_after_gauge} can be solved by
the method of Green-functions. Such GF's have been found
\cite{LandHor1991,OronHorwitz2001} through a 5-fold Fourier
transform of the wave equation. The GF obeys the wave-equation of a
point source
\begin{align}
    \label{eq:hps_green_function_definition}
    \partial_{\beta} \partial^{\beta} g(x,\tau) = \delta^4(x) \delta(\tau)
\end{align}
After transformation to momentum space,
\eqref{eq:hps_green_function_definition} becomes
($k^2 = k_{\mu}k^{\mu}$)
\begin{align}
    \label{eq:hps_green_function_fourier_Transform}
    (k^2 + \sigma_5 k_5^2) \tilde{g}(k,k_5) = 1
\end{align}
i.e., in terms of the inverse transform
\begin{align}
    \label{eq:hps_green_function_Fourier_solution}
    g(x,\tau) =
        \dfrac{1}{(2\pi)^5}
        \int d^4k \, dk_5
            \dfrac{1}{k^2 + \sigma_5 k_5^2}
            e^{i[k \cdot x + \sigma_5 k_5 \tau]}
\end{align}
\\

Although \eqref{eq:hps_green_function_Fourier_solution} is not a
well defined integral, there are, as for GF's in 4D, several ways of
defining the integral, which result in GF's all of which satisfy
\eqref{eq:hps_green_function_definition}. These different forms of
solutions have physical consequences and it is part of the
motivation of our work to obtain criteria which could determine this
choice.

For example, Land and Horwitz \cite{LandHor1991} found what
they called the \emph{Principal-Part} GF to be
\begin{align}
    g_{P}(x,\tau) & =
        - \dfrac{1}{4\pi  } \delta(x^2) \delta(\tau)
        - \dfrac{1}{2\pi^2}
          \dfrac{\partial}{\partial x^2}
            \dfrac{\theta (-\sigma x^2 - \tau^2)}{\sqrt{-\sigma x^2 - \tau^2}}
\end{align}
where $\sigma = \sigma_5 = \pm 1$ is the signature of the fifth dimension, $\tau$.

A later work by Oron and Horwitz \cite{OronHorwitz2001} found
another, $\tau$-retarded GF of the form
\begin{align}
    g(x,\tau) & =
        \dfrac{2 \theta(\tau)}{(2 \pi)^3}
        \begin{cases}
            \dfrac{1}{(-x^2 - \tau^2)^{3/2}}
                \tan^{-1} \left( \dfrac{1}{\tau} \sqrt{-x^2 - \tau^2}\right)
            - \dfrac{\tau}{x^2 (x^2 + \tau^2)}
            \qquad
            &
            x^2 + \tau^2 < 0                    \\
            \\
            \dfrac{1}{2}
            \dfrac{1}{(x^2 + \tau^2)^{3/2}}
            \ln
                \Bigg|
                    \dfrac{\tau - \sqrt{\tau^2 + x^2}}
                          {\tau + \sqrt{\tau^2 + x^2}}
                \Bigg|
                - \dfrac{\tau}{x^2 (x^2 + \tau^2)}
            &
            x^2 + \tau^2 > 0
        \end{cases}
\end{align}
where only the $(4,1)$ case was explicitly given.

GF's of $(n,1)$ wave equations are well known in the mathematical
literature (this is taken from \cite{Kythe1996}, cf. also
\cite{Gelfand1964_1}, \cite{Kazinski2002}, \cite{GalTsov2002}):
\begin{align}
    g(x,t)
    & =
        \begin{cases}
                \dfrac{\theta(t)}{2\pi}
                \left(
                    \dfrac{1}{\pi}
                    \dfrac{d}{d t^2}
                \right)^{(n-3)/2}
                \delta(t^2 - \vb{x}^2)
                &
                n = 3, 5, 7, \cdots     \\
            \\
                \dfrac{1}{2\pi}
                \left(
                    \dfrac{1}{\pi}
                    \dfrac{d}{d t^2}
                \right)^{(n-2)/2}
                \dfrac{\theta(t - |\vb{x}|)}
                      {\sqrt{t^2 - \vb{x}^2}}
                &
                n = 4, 6, 8, \cdots
            \\
        \end{cases}
\end{align}
where $n=4$ is the case of $(4,1)$ metric.
These well known GF's, which are \emph{retarded} in $t$, can be made
\emph{symmetric} by proper choice of
contour of integration on the original Fourier integral representation.

Once the GF's have been found,
the general field generated by a given source can then be found by
integration on its support
\begin{align}
    \label{eq:hps_potential_solution_using_green_function}
    a^{\alpha}(x,\tau) =
            e_0
            \int d^4x' \, d\tau'
                \,
                g(x-x',\tau-\tau') \, j^{\alpha}(x',\tau')
\end{align}
and applying it on a point particle given by \eqref{eq:ose_current_of_point_event}.
The potentials of point events can then be found
\begin{align}
    \label{eq:hps_potential_solution_using_green_function_for_point_particle}
    a^{\alpha}(x,\tau) =
            e_0
            \int_{-\infty}^{+\infty}
                d\tau'
                \,
                g(x-z(\tau'),\tau-\tau') \, \dot{z}^{\alpha}(\tau')
\end{align}
\\

In order to get some insight into the criteria for choosing
appropriate GF's, we study solutions of the differential equation
\eqref{eq:ose_potentials_wave_equation_after_gauge} (for the
particular choice of uniformly moving point source) without using
the GF, i.e., we compute directly
%%%%
\begin{align}
    \label{eq:ose_potential_solution_with_green_function_point_particle}
    a^{\alpha}(x,\tau) & =
            \dfrac{e}{(2\pi)^5}
            \int_{-\infty}^{+\infty} d\tau' \,
                \dot{z}^{\alpha}(\tau')
                \int d^4k \, dk_5 \,
                    \dfrac{e^{i[k \cdot (x - z(\tau')) + \sigma_5 k_5 (\tau - \tau') ]}}
                          {k^2 + \sigma_5 k_5^2}
\end{align}
\\

Solutions  of the integral
\eqref{eq:ose_potential_solution_with_green_function_point_particle}
are the subject of the next section.
%%
%%
%%
%%
%% END OF INTRODUCTION
%%
%%
%%

\section{Fields of a uniformly moving point charge}
%%%    \input{UMP_Fields}
% =============================================================== %
% --------------------------------------------------------------- %
% ----------------------> Solutions <---------------------------- %
% --------------------------------------------------------------- %
% =============================================================== %
\label{sec:solutions}

\subsection{Solutions of the wave equation}
Let us seek a solution to the field equation generated by a
uniformly moving point source. Such a source has a general worldline
description
\begin{align}
    \label{eq:ums_worldline}
    z^{\alpha}(\tau) & = z^{\alpha}_0 + b^{\alpha} (\tau - \tau_0) \equiv
                         D^{\alpha}   + b^{\alpha} \tau
    \qquad
    \qquad
    \dot{z}^{\alpha}(\tau) = b^{\alpha}
    \qquad
    \alpha \in \left\{ 0,1,2,3,5 \right\}
\end{align}
where for $z^{5} \equiv \tau$ we have $b^{5} = 1$. However, we leave
$b^{5}$ unspecified, leaving the possibility for a 5D symmetry of
the solution to emerge, as indeed we find. Without loss of
generality, we can eliminate $D^{\alpha}$ by choosing a coordinate
system in which $D^{\alpha} = 0$. The current of such a source is
then given by
\begin{align}
    \label{eq:def_current_of_uniformly_moving_event}
    j^{\alpha}(x,\tau) & =
        b^{\alpha} \delta^{4} \left[
                                x - b \tau
                              \right]
\end{align}
\\

Substituting \eqref{eq:ums_worldline} into
\eqref{eq:ose_potential_solution_with_green_function_point_particle}
we obtain an integral representation of the uniform motion fields,
which could be called \emph{pre-Coulomb fields}:
\begin{align}
    a^{\alpha}(x,\tau) & =
            \dfrac{e}{(2\pi)^5}
            \int_{-\infty}^{+\infty} d\tau' \,
                b^{\alpha}
                \int d^4k \, dk_5 \,
                    \dfrac{e^{i[k \cdot (x - b \tau') + \sigma_{5} k_5 (\tau - b^{5} \tau') ]}}
                          {k^2 + \sigma_{5} k_5^2} =
    \nonumber \\
    & =
            \dfrac{e \, b^{\alpha}}{(2\pi)^5}
            \int_{-\infty}^{+\infty} d\tau' \,
                \int d^5k \,
                    \dfrac{e^{i k_{\beta}[x^{\beta} - b^{\alpha} \tau']}}
                          {k_{\beta} k^{\beta}} =
    \nonumber \\
    & =
            \dfrac{e \, b^{\alpha}}{(2\pi)^5}
            \int d^5k \,
                \int_{-\infty}^{+\infty} d\tau' \,
                    \dfrac{e^{i k_{\beta}[x^{\beta} - b^{\alpha} \tau']}}
                          {k_{\beta} k^{\beta}} =
    \nonumber \\
    & =
            \dfrac{e \, b^{\alpha}}{(2\pi)^5}
            \int d^5k \,
                    \dfrac{e^{i k_{\beta}x^{\beta}}}
                          {k_{\beta} k^{\beta}}
                \int_{-\infty}^{+\infty} d\tau' \,
                    e^{- i k_{\beta}b^{\alpha} \tau'} =
    \nonumber \\
    & =
            \dfrac{e \, b^{\alpha}}{(2\pi)^4}
            \int d^5k \,
                    \dfrac{e^{i k_{\beta}x^{\beta}}}
                          {k_{\beta} k^{\beta}}
                    \,
                    \delta
                    \left(
                        k_{\beta} b^{\beta}
                    \right)
\end{align}

The argument $k_{\beta} b^{\beta}$ of the $\delta$-function causes
the 5-fold integration to be constrained to a 5D hyperplane,
\begin{align*}
    S[b] & = \left\{
                k \in \mathbb{R}^5
                \big|
                k_{\alpha} b^{\alpha} = 0
             \right\}
\end{align*}
whose normal is just $b^{\beta}$.

In order to proceed, we select a pivot axis, for which integration
would put the remaining 4-fold integral to be in that hyperplane.
Naturally, we select $k^{5}$, since we will take $b^{5} > 0$:
\begin{align}
    \begin{split}
        k_{\beta} b^{\beta} & =
            k \cdot b + \sigma_{5} k^{5} b^{5} =
            b^{5}
            \left[
                k \cdot b' + \sigma_{5} k^{5}
            \right] \\
        \intertext{where $\mu \in \{ 0,1,2,3 \}$ and}
        b'^{\mu}
        & =
            \dfrac{b^{\mu}}{b^{5}}
        \\
        \intertext{$b'^{\mu}$ is the \emph{$(3,1)$ velocity} of the source relative
                   to its motion in the $\tau$ direction.}
        \intertext{We then have}
        \delta \left( k_{\beta} b^{\beta} \right) & =
            \dfrac{1}{|b^{5}|}
            \delta
                \left(
                    k \cdot b' + \sigma_{5} k^{5}
                \right)
            \Longrightarrow
            k^{5} = - \dfrac{1}{\sigma_{5}} k \cdot b'
        \\
        k_{\beta} k^{\beta} & =
            k^2 + \sigma_{5} (k^{5})^2 =
            k^2 + \sigma_{5} \left[
                            \dfrac{-1}{\sigma_{5}}
                            \left(
                                k \cdot b'
                            \right)
                         \right]^2
            =
            k^2 + \sigma_{5} (k \cdot b')^2
        \\
        k_{\beta} x^{\beta} & =
            k \cdot x + \sigma_{5} k^{5} \tau =
            k \cdot x + \sigma_{5} \dfrac{-1}{\sigma_{5}}
                               \left[
                                    k \cdot b'
                               \right]
                               \tau =
            k \cdot (x - b' \tau)
    \end{split}
\end{align}
And thus, we obtain:
\begin{align}
    \label{eq:a_field_fourier_4_fold_integral}
    \begin{split}
        a^{\alpha}(x,\tau) & =
            \dfrac{e \, b^{\alpha}}{2 \pi^4 |b^{5}|}
            \int d^4k \,
                    \dfrac{e^{i k \cdot (x - b'\tau)}}
                          {k^2 + \sigma_{5} (k \cdot b')^2}
        \\
    \end{split}
\end{align}
The integral \eqref{eq:a_field_fourier_4_fold_integral} can be
solved by introducing a rotation in $k$ space in which $b'$ takes a
particularly simple form, namely, along one of the axes. Aside from
the special case of $b'^2 = 0$, $b'$ can be rotated to be along one
of the axes by an $SO(3,1)$ transformation. We shall now divide our
discussion to the \emph{spacelike} case where $b'^2 > 0$ and the
\emph{timelike} case where $b'^2 < 0$, and to avoid complications,
we shall solve the zero measure case of $b'^2 = 0$ by a limiting
procedure.

\subsection{$a$-fields due to a $(3,1)$ timelike source}
Since $b'^2 < 0$, we can find $\Lambda \in SO(3,1)$ such that
\begin{align*}
    b'     & = \Lambda b''  \qquad \text{such that} \qquad b'' =  \left[ b''^0 ; \vb{0} \right]  \\
    b''^0  & \equiv s = \epsilon(b'^0) \sqrt{-b'^2}             \\
\end{align*}
and since $|\Lambda| = 1$, we have $d^4 k'' = d^4k$. Replacing $b'$
with $b''$ and $k$ with $k''$, we obtain:
\begin{align}
    \label{eq:a_field_fourier_4_fold_integral_diagonalized}
    a^{\alpha}(x,\tau) & =
        \dfrac{e \, b^{\alpha}}{2 \pi^4 |b^{5}|}
        \int d^4k'' \,
                \dfrac{e^{i \left[ \vb{k}'' \cdot \vb{x}''
                                    - k''^{0} (x''^0 - s\tau)
                            \right]
                         }
                      }
                      {k''^2 + \sigma_{5} (- k''^0 s)^2} =
    \nonumber
    \\
    & =
        \dfrac{e \, b^{\alpha}}{2 \pi^4 |b^{5}|}
        \int d^4k'' \,
                \dfrac{e^{i \left[ \vb{k}'' \cdot \vb{x}''
                                    - k''^{0} (x''^0 - s\tau)
                            \right]
                         }
                      }
                      {\vb{k}''^2  + (k''^{0})^2 (\sigma_{5} s^2 - 1)}
\end{align}
where $x'' = \Lambda^{-1} x$. We follow the convention of boldface
corresponding to the space part of a four-vector. Since  $b'' \tau$
is a 4-vector along the time axis, we can find simple closed form
expressions for $x''$ as follows:
\begin{align*}
    x''^{0} & = \dfrac{x''^{0} s}{s} =
                \dfrac{ - x'' \cdot b''}{\sqrt{-b'^2}} =
                \dfrac{ - x \cdot b'}{\sqrt{-b'^2}}             \\
    (\vb{x}''^{0})^2
            & = x''^2 + (x''^0)^2 =
                x^2   + \dfrac{ (b' \cdot x)^2}{ - b'^2} =
                x^2   - \dfrac{ (b' \cdot x)^2}{ b'^2 }
\end{align*}
Integral \eqref{eq:a_field_fourier_4_fold_integral_diagonalized}
depends on the value of the denominator along the path of
integration (\ref{table:regions_of_source_motion}), where 2 types of
source motion emerge. The types are given as follows:

\begin{center}
    \renewcommand{\arraystretch}{1.2}
    \begin{tabular}{|c|p{.8\linewidth}|}
        \hline
        Source motion        &   Description                     \tabularnewline[8pt]
        \hline
        \hline
        $\sigma_{5} b'^2 > 1 $  &  \emph{Supershell} case, where the integral
                                   \eqref{eq:a_field_fourier_4_fold_integral_diagonalized}
                                   has a well defined solution,
                                   essentially the Laplace GF in 4D.    \\
        \hline
        $\sigma_{5} b'^2 < 1$   &  \emph{Undershell} case, where the integral
                                   is not well defined.
                                   The integral is essentially the Maxwell GF.          \\
        \hline
    \end{tabular}
    \label{table:regions_of_source_motion}
    %%\caption{Regions of source motion}
\end{center}

In the following we describe the properties of these cases.

\subsubsection{Undershell timelike $a$-fields $\sigma_{5} b'^2 > -1$}
As mentioned, the denominator of integral
\eqref{eq:a_field_fourier_4_fold_integral_diagonalized} is not
positive definite. Nevertheless, we shall proceed first by absorbing
the coefficient $1 - \sigma_{5} s^2$ into $k^{0}$:
\begin{align*}
    k^{0} \longrightarrow \dfrac{k^0}{\sqrt{1 - \sigma_{5} s^2}}
\end{align*}
Equation \eqref{eq:a_field_fourier_4_fold_integral_diagonalized}
obtains the form
\begin{align*}
    a^{\alpha}(x,\tau) & =
        \dfrac{e \, b^{\alpha}}{2 \pi^4 |b^{5}| \sqrt{1 - \sigma_{5} s^2}}
        \int d^4k \,
                \dfrac{e^{i \left[ \vb{k} \cdot \vb{x}''
                                    - k^{0}  \dfrac{(x''^0 - s\tau)}{\sqrt{1 - \sigma_{5} s^2}}
                            \right]
                         }
                      }
                      {\vb{k}^2  - (k^{0})^2} =
    \\
    & =
        \dfrac{e \, b^{\alpha}}{|b^{5}| \sqrt{1 - \sigma_{5} s^2}}
        \left\{
            \dfrac{1}{(2 \pi^4)}
            \int d^4k \,
                    \dfrac{e^{i k \cdot y}}
                          {k^2}
        \right\}
    \\
    \intertext{where}
    y^{\mu} & \equiv
            \left[
                \dfrac{x''^0 - s \tau}{\sqrt{1 - \sigma_{5} s^2}}
                ;
                \vb{x}''
            \right]
\end{align*}
The last integral inside the braces is the well known Fourier
integral form for Maxwell wave-equations's GF in 4 dimensions, and
though it is ill-defined, it has many well known solutions,
corresponding to different limits of the integration contour chosen.
We shall choose the \emph{Principal Part} solution for our present
study.
\begin{align}
    \label{eq:Maxwell_Green_Function_Principal_Part}
    G_{P}(x) & =
        \dfrac{1}{(2\pi)^4}
        P
        \int_{\mathbb{R}^4} d^4 k \,
            \dfrac{e^{i k \cdot x}}{k^2} =
        \dfrac{\delta(x^2)}{4 \pi}
\end{align}
corresponding the sum of retarded and advanced GF's. Using the
$G_{P}$ above, we arrive at the undershell $a$-fields:
\begin{align}
    \label{eq:a_fields_solution_b_timelike_below_shell}
    a^{\alpha}(x,\tau) & =
        \dfrac{e \, b^{\alpha}}{4 \pi |b^{5}| \sqrt{1 - \sigma_{5} s^2}}
        \delta
            \left[
                \vb{x}''^2
                -
                \dfrac{(x''^0 - s\tau)^2}{1 - \sigma_{5} s^2}
            \right] =
    \nonumber
    \\
    & =
        \dfrac{e \, b^{\alpha}}{4 \pi |b^{5}|  \sqrt{1 + \sigma_{5} b'^2}}
        \delta
            \left[
                x^2
                -
                \dfrac{(b' \cdot x)^2}{b'^2}
                +
                \dfrac{(-b' \cdot x + b'^2 \tau)^2}{b'^2 (1 + \sigma_{5} b'^2)}
            \right]
\end{align}
We call these \emph{undershell solutions} because they correspond to
the offshell mass of the source below its Galilean target mass.
Equation \eqref{eq:a_fields_solution_b_timelike_below_shell} has
$O(3,1)$ symmetry, with respect to the lower 4 coordinates of
$x^{\alpha}$. It can be further broken to a sum of
$\delta$-functions with linear arguments in $\tau$, as follows:
\begin{align*}
    p(\tau_{1,2}) & \equiv
                x^2
                -
                \dfrac{(b' \cdot x)^2}{b'^2}
                +
                \dfrac{(-b' \cdot x + b'^2 \tau_{1,2})^2}{b'^2 (1 + \sigma_{5} b'^2)} = 0 \\
    \tau_{1,2}  & =
                \dfrac{b' \cdot x}{b'^2}
                \pm
                \dfrac{1}{b'^2}
                \sqrt{1 + \sigma_{5} b'^2}
                \sqrt{(b' \cdot x)^2 - b'^2 x^2} \\
    \intertext{Using the linearity of the $\delta$-function}
    \delta ( p(\tau) ) & =
                \Bigg| \dfrac{b'^2 (1 + \sigma_{5} b'^2)}{(b'^2)^2} \Bigg|
                \delta \left[
                            (\tau - \tau_1) (\tau - \tau_2)
                       \right] =
                \dfrac{(1 + \sigma_{5} b'^2)}{b'^2 |\tau_1 - \tau_2|}
                \left[
                    \delta(\tau - \tau_1)
                    +
                    \delta(\tau - \tau_2)
                \right] \\
    \intertext{we then have}
    a^{\alpha}(x,\tau) & =
        \dfrac{e \, b^{\alpha} \, \Delta_{+}}
              {8 \pi |b^{5}|  \sqrt{(b' \cdot x)^2 - b'^2 x^2}} \\
    \intertext{where}
    \Delta_{+}(x,\tau) & = \delta(\tau - \tau_1) + \delta(\tau - \tau_2)
\end{align*}
The $a$-field depends on the the fifth metric component,
$\sigma_{5}$, only through $\tau_{1,2}$, i.e., the coefficient is
independent of the signature of the 5D space. However, the values
$\tau = \tau_{1,2}$ correspond to a 4D surface in 5D space where the
$a$-fields are non-zero, and therefore, the metric appears, to some
extent, in the \emph{geometry} of the non-zero surfaces.

After some algebra, the $a$-fields can gain yet another, 5D
covariant form. We start by rewriting the $\delta$-function argument
in 5D form:
\begin{align}
    \label{eq:p_tau_5d_form}
    p(\tau) & =
                x^2
                -
                \dfrac{(b' \cdot x)^2}{b'^2}
                +
                \dfrac{(-b' \cdot x + b'^2 \tau_{1,2})^2}{b'^2 (1 + \sigma_{5} b'^2)} =   \nonumber \\
            & =
                x_{\beta} x^{\beta} -
                \dfrac{b_{\beta} x^{\beta}}{b_{\beta} b^{\beta}}
\end{align}
where the metric signature of $(4,1)$ or $(3,2)$ is used in the
contraction products, e.g.:
\begin{align*}
    b_{\beta} x^{\beta} & =
        b\cdot x + \sigma_{5} b^{5} x^{5} = b \cdot x + \sigma_{5} b^{5} \tau
\end{align*}
For $(4,1)$ case, we have:
\begin{align*}
    b_{\alpha} b^{\alpha} & = b^2 + (b^{5})^2 = \dfrac{1}{(b^{5})^2} \left[b'^2 + 1 \right] > 0 \\
\end{align*}
since $\sigma_{5} b^2 > -1$ in the present case. Thus, in the
$(4,1)$ metric, the \emph{undershell} motion corresponds to the 5D
\emph{spacelike} region in the $b^{\alpha}$ velocity space. As shall
be observed later, this region is not limited to 4D timelike source
motion $b'^2 < 0$, and it includes $b'^2 \geq 0$ as well. For the
$(3,2)$ metric, on the other hand, the motion is in the 5D timelike
region $b_{\alpha} b^{\alpha} < 0$.

Furthermore, one can define
\begin{align}
    \label{eq:n_5d_definition}
    n^{\alpha} & =
        \dfrac{b^{\alpha}}{\sqrt{\sigma_{5} b_{\beta} b^{\beta}}}
\end{align}
Substituting equations \eqref{eq:p_tau_5d_form}
,\eqref{eq:n_5d_definition} into
\eqref{eq:a_fields_solution_b_timelike_below_shell}, we arrive at
the 5D covariant form:
\begin{align}
    a^{\alpha}(x,\tau) & =
        \dfrac{e n^{\alpha}}{4 \pi}
        \delta
            \left[
                x_{\beta} x^{\beta}
                -
                \sigma_{5} (n_{\beta} x^{\beta})^2
            \right]
\end{align}

The term \emph{undershell source motion} stems from the mass shell
equation
\begin{align}
    P_{\alpha} P^{\alpha} & = M^2 \dot{x}_{\alpha} \dot{x}^{\alpha} =
                              M^2 \left[
                                    -\dfrac{m^2}{M^2} + \sigma_{5}
                                  \right] =
                              \sigma_{5} M^2
                              \left[
                                - \sigma_{5} \dfrac{m^2}{M^2}
                                +
                                1
                              \right] = \\
                          & =
                              M^2 b_{\alpha} b^{\alpha} \\
    \intertext{or}
    b_{\alpha} b^{\alpha}  & = \sigma_{5} \left[ - \sigma_{5} \dfrac{m^2}{M^2} + 1 \right]
\end{align}

For $(4,1)$ metric, $\sigma_{5} = 1$, where undershell timelike
motion leads to $|m| < M$. Hence, undershell motion refers to the
\emph{mass-shell of the source} being \emph{less} than its
non-relativistic mass-shell $M$.

\subsubsection{Supershell timelike $a$-fields $\sigma_{5} b'^2 < -1$}
As the name suggests, in the $(4,1)$ metric of the source motion,
the \emph{supershell} case is determined by $|m| > M$, i.e., the
relativistic mass $|m|$ being greater than the its non-relativistic
Galilean target mass $M$. In this case, however, only $\sigma_{5} =
1$ is applicable, since there is no timelike motion $b'^2 < 0$ such
that $(-1)b'^2 + 1 < 0$, unless $b'^2 > 1$, which is no longer
timelike. Such a case will be investigated later. In this case, the
integral \eqref{eq:a_field_fourier_4_fold_integral_diagonalized} is
well defined, as the zeros in the denominator are no longer real. By
following a similar procedure of absorbing the coefficient of $k^0$
in the denominator into $k^0$, we obtain ($\sigma_{5} = 1$):
\begin{align*}
    a^{\alpha}(x,\tau) & =
        \dfrac{e \, b^{\alpha}}{2 \pi^4 |b^{5}| \sqrt{s^2 - 1}}
        \int d^4k \,
                \dfrac{e^{i \left[ \vb{k} \cdot \vb{x}''
                                    - k^{0}  (x''^0 - s\tau)/{\sqrt{s^2 - 1}}
                            \right]
                         }
                      }
                      {\vb{k}^2  + (k^{0})^2} =
    \\
    & =
        \dfrac{4\pi e \, b^{\alpha}}{(2 \pi)^4 |b^{5}| \, |\vb{x''}| \sqrt{s^2 - 1}}
        \int_{0}^{+\infty}  k \,  dk \,
            \sin (k |\vb{x''}|)
            \int_{-\infty}^{+\infty} dk^0 \,
                \dfrac{e^{-i k^{0} [x''^{0} - s\tau]/{\sqrt{s^2 - 1}}}}
                      {k^2 + (k^{0})^2}
                 =
    \\
    & =
        \dfrac{ e \, b^{\alpha}}{4 \pi^3 |b^{5}| \, |\vb{x''}| \sqrt{s^2 - 1}}
        \int_{0}^{+\infty}  k dk \,
            \sin (k |\vb{x''}|)
            (-1) \dfrac{\pi}{k} e^{-k [x''^{0} - s\tau]/{\sqrt{s^2 - 1}}|} =
    \\
    & =
        \dfrac{e \, b^{\alpha}}{4 \pi^2 |b^{5}| \, |\vb{x''}|  \sqrt{s^2 - 1}}
        \dfrac{1}{2i}
        \left[
            \dfrac{0 - 1}{- |x''^{0} - s\tau|/{\sqrt{s^2 - 1}} + i|\vb{x''}|}
            -
            \dfrac{0 - 1}{- |x''^{0} - s\tau|/{\sqrt{s^2 - 1}} - i|\vb{x''}|}
        \right] =
    \\
    & =
        \dfrac{e \, b^{\alpha} }
              {4 \pi^2 |b^{5}| \sqrt{-b'^2 - 1}}
        \dfrac{1}
              {
                \left[
                    \vb{x''}^2
                    +
                    \dfrac{(x''^{0} - s\tau)^2}{-b'^2 - 1}
                \right]
              } =
    \\
    & =
        \dfrac{e \, b^{\alpha}}
              {4 \pi^2 |b^{5}| \sqrt{-b'^2 - 1}}
        \dfrac{1}
              {
                \left[
                    x^2 - \dfrac{(b' \cdot x)^2}{b'^2}
                    +
                    \dfrac{(-b' \cdot x + b'^2 \tau)^2}{b'^2 (b'^2 + 1)}
                \right]
              } =
    \\
    & =
        \dfrac{e \, n^{\alpha}}
              {4 \pi^2}
        \dfrac{1}
              {
                \left[
                    x_{\beta} x^{\beta} + (n_{\beta} x^{\beta})^2
                \right]
              }
\end{align*}
where we have defined $n^{\alpha} \equiv \dfrac{b^{\alpha}}{\sqrt{ -
b_{\beta} b^{\beta}}}$.

The supershell $a$-field is found to be a \emph{smooth} function on
5D spacetime; it is the 5D analogue of the well-known (cf.
\cite{Jackson1995}) on-shell 4D Maxwell $a$-field of a uniformly
moving charge

\begin{align*}
    A^{\mu}(x) & = \dfrac{e n^{\mu}}{4 \pi \sqrt{x^2 + (n \cdot x)^2}}
\end{align*}
where $n^{\mu} = \dfrac{d z^{\mu}(s)}{ds}$ is the constant
4-velocity obeying the mass-shell constraint $n^2 = -1$. Clearly,
the 5D $a$-field, proportional to $r^{-2}$, as opposed to the
Maxwell $A$-field being proportional to $r^{-1}$, is a consequence
of the additional dimension.

The supershell $a$-field has the same form as the GF for the $4D$
\emph{Laplace operator}:
\begin{align}
    \nabla^2 G_{L}(x) & = \delta^4(x)
    \\
    \intertext{where
        $x \in \mathbb{R}^4$,
        $r^2 = (x^{1})^2 + (x^{2})^2 + (x^{3})^2 + (x^{4})^2$
    } \nonumber
\end{align}
and the GF $G_{L}(x)$ is given by (cf. \cite{Kythe1996}):
\begin{align}
    G_{L}
    & =
        \dfrac{1}{(2\pi)^4}
        \int
            \dfrac{e^{i k_{c} x^{c}}}{k_{c} k^{c}}
    =
        \dfrac{1}{2\pi^2 r^2}
\end{align}
where in this case
\begin{align*}
    k_{c} x^{c} & = k^{1} x^{1} + k^{2} x^{2} + k^{3} x^{3} + k^{4} x^{4}
\end{align*}

This is far from being coincidental, as the 5D scalar $x_{\beta}
x^{\beta} + (n_{\beta} x^{\beta})^2$ reduces to $r^2$ in 4D, when
the source's uniform velocity is purely \emph{temporal} $n^{\beta} =
[1;0,0,0,0]$.

\subsection{$a$-fields due to a spacelike source}
We now solve the integral \eqref{eq:a_field_fourier_4_fold_integral}
for spacelike source motion, $b'^2 > 0$, which may not be regarded
as a possible physical source, since it implies faster than light
motion of the source particle. Once again, we choose to integrate in
a $k$-frame such that $b'' = [0; 0,0,b''^{3}]$ is along one of the
spatial axes, e.g., the $z$-axis, and we now define
\begin{align}
    b''^3 & \equiv s = \sqrt{b''^2} = \sqrt{b'^2}, \qquad b'^2 > 0
\end{align}

The current in the $b''$ frame can be expressed by
\begin{align}
    \label{eq:local_frame_spacelike_current}
    j''^{\mu}(x'',\tau) & = b''^{\mu} \delta^4 [x - z''(\tau)] =
                        b'^{3} \delta^{\mu}_{3} \delta(''t) \delta(x'') \delta(y'') \delta(z'' - b''^{3} \tau)
\end{align}

Returning to pre-Maxwell $a$-field integral
\eqref{eq:a_field_fourier_4_fold_integral},
\begin{align}
    \label{eq:a_field_fourier_4_fold_integral_spacelike}
    a^{\alpha}(x,\tau) & =
        \dfrac{e \, b^{\alpha} }{|b^{5}| (2 \pi)^4}
        \int d^4 k''
            \dfrac{e^{i [k''^1 x'' + k''^2 y'' + k''^3 (z'' - s\tau) - k''^0 t'']}}
                  {(k''^1)^2 + (k''^2)^2 + (k''^3)^2 - (k''^{0})^2 + \sigma_{5}(s k^3)^2}
        =
    \\
    & =
        \dfrac{e \, b'^{\alpha} }{|b^{5}| (2 \pi)^4}
        \int d^4 k
            \dfrac{e^{i [k^1 x + k^2 y + k^3 (z - s\tau) - k^0 t]}}
                  {(k^1)^2 + (k^2)^2 + (k^3)^2 (1 + \sigma_{5} s^2) - (k^{0})^2}
\end{align}
where we have renamed $k''$ as $k$ and $x''$ as $x$.

The coefficient of $(k^{3})^2$ changes sign when $1 + \sigma_{5} s^2
 = 1 + \sigma b'^2 = 0$ which can only occur when $\sigma_{5} = -1$ (since for spacelike motion, $b'^2 > 0$)
and $|s| \geq 1$ ($s = \epsilon(b'^{0}) \sqrt{b'^2}$), causing the
denominator to obtain a $(2,2)$ quadratic form.

Once again, the form of the fields are characterized by the types of
source motion. In the following, we shall treat both possible types
of source motion separately.

\subsubsection{Under Spacelike motion $1 + \sigma_{5} b'^2 > 0$}
Rescaling $k^{3} \rightarrow k^{3} / \sqrt{1 + \sigma_{5} s^2}$, the
spacelike $a$-field integral
\eqref{eq:a_field_fourier_4_fold_integral_spacelike} can be
expressed by
\begin{align*}
    a^{\alpha}(x,\tau) & =
        \dfrac{e \, b^{\alpha} }{|b^{5}| (2 \pi)^4 \sqrt{1 + \sigma_{5} s^2}}
        \int d^4 k
            \dfrac{e^{i k \cdot y}}
                  {k^2} = \\
    & =
        \dfrac{e \, b^{\alpha} \, \delta(y^2)}{4 \pi |b^{5}| \sqrt{1 + \sigma_{5} s^2}}
\end{align*}
where $y^{\mu} = \left[ x^0 ; x^1, x^2, \dfrac{x^3 - s\tau}{\sqrt{1+\sigma_{5} s^2}} \right], k^2 = k_{\mu} k^{\mu}$,
and we have chosen the \emph{Principal Part} contour.

However:
\begin{align*}
    y^2 & = (x^1)^2 + (x^2)^2 + \dfrac{(x^3 - s\tau)^2}{1 + \sigma_{5} s^2} - (x^0)^2 =           \\
        & = x_{\mu} x^{\mu} - (x^3)^2 + \dfrac{(x^3 - \sqrt{b'^2} \tau)^2}{1 + \sigma_{5} b'^2} \\
    \intertext{We can furthermore put $x^{3}$ into an invariant form:}
    x^3 & = \dfrac{x^3 \, s }{s} = \dfrac{x \cdot b'}{\sqrt{b'^2}}              \\
    y^2 & =     x^2
            -
                \dfrac{(b' \cdot x)^2}{b'^2}
            +
                \dfrac{(-b' \cdot x + b'^2 \tau)^2}
                      {b'^2 (1 + \sigma_{5} b'^2)}
\end{align*}
The spacelike $a$-fields then obtain a 4D covariant form:
\begin{align*}
    a^{\alpha}(x,\tau) & =
        \dfrac{e \, b^{\alpha}}{4 \pi |b^{5}| \sqrt{1 + \sigma_{5} b'^2}}
        \,
        \delta
            \left[
                    x^2
                -
                    \dfrac{(b' \cdot x)^2}{b'^2}
                +
                    \dfrac{(-b' \cdot x + b'^2 \tau)^2}
                        {b'^2 (1 + \sigma_{5} b'^2)}
            \right] = \\
    & =
        \dfrac{e \, b^{\alpha} \, \Delta_{+}}{8 \pi |b^{5}| \sqrt{(b' \cdot x)^2 - b'^2 x^2}}    \\
    \intertext{where we have, as before:}
    \Delta_{+} & \equiv \delta(\tau - \tau_1) + \delta(\tau - \tau_2)   \\
    \tau_{1,2}   & =      \dfrac{b' \cdot x}{b'^2}
                        \pm
                          \dfrac{1}{b'^2}
                          \sqrt{1 + \sigma_{5} b'^2}
                          \sqrt{(b' \cdot x)^2 + b'^2 x^2}
\end{align*}
Thus, the \emph{under} spacelike motion fields are of the same form
as their timelike \emph{under-shell} counterparts.

\subsubsection{Super Spacelike motion $1 + \sigma_{5} b'^2 > 0$}
As mentioned above, in this case we have $b'^2 > 1$ and the choice
$\sigma_{5} = -1$ is therefore necessary. Therefore, the integral
\eqref{eq:a_field_fourier_4_fold_integral_spacelike} takes the form:
\begin{align}
    a^{\alpha}(x,\tau) & =
        \dfrac{e \, b^{\alpha}}{|b^{5}| (2 \pi)^4}
        \int d^4 k
            \dfrac{e^{i [k^1 x + k^2 y + k^3 (z - s\tau) - k^0 t]}}
                  {(k^1)^2 + (k^2)^2 - (s^2 - 1) (k^3)^2 - (k^{0})^2}
    \nonumber \\
    \intertext{Rescaling $k^{3} \rightarrow k^3 / \sqrt{s^2 - 1}$:}
    & =
        \label{eq:a_field_4_fold_integration_super_space_like_intermediate_1}
        \dfrac{e \, b^{\alpha}}{(2 \pi)^4 |b^{5}| \sqrt{s^2 - 1}}
        \int d^4 k
            \dfrac{e^{i k \cdot y}}
                  {(k^1)^2 + (k^2)^2 - (k^3)^2 - (k^{0})^2}
\end{align}
We shall solve this integral by transforming the integrand to a
Gaussian form
\begin{align*}
    \int_{-\infty}^{+\infty}
        \dfrac{e^{iax}}{x^2}
        dx
        =
        \dfrac{1}{2i}
        \int_{-\infty}^{+\infty}
            \varepsilon(q)
            \,
            dq
            \,
            \int_{-\infty}^{+\infty}
                e^{i a x + i q x^2}
                dx
\end{align*}
where $\varepsilon(q)$ is the sign function. Using this relation in
\eqref{eq:a_field_4_fold_integration_super_space_like_intermediate_1}
we obtain
\begin{align*}
    a^{\alpha}(x,\tau) & =
        \dfrac{e \, b^{\alpha}}{(2 \pi)^4 |b^{5}| \sqrt{s^2 - 1}}
        \dfrac{1}{2i}
        \int_{-\infty}^{+\infty} dq \,
            \varepsilon(q)
            \int d^4 k
                \exp
                    \left[
                        i k \cdot y
                        +
                        i q [(k^1)^2 + (k^2)^2 - (k^3)^2 - (k^{0})^2]
                    \right]
        =
    \\
    & =
        \dfrac{e \, b^{\alpha}}{(2 \pi)^4 |b^{5}| \sqrt{s^2 - 1}}
        \dfrac{1}{2i}
        \int_{-\infty}^{+\infty} dq \,
            \varepsilon(q)
            \left(
                \sqrt{\dfrac{\pi}{q}}
            \right)^4
            \exp
                \left[
                    \dfrac{i}{4q}
                    \left(
                        (y^1)^2 + (y^2)^2 - (y^3)^2 - (y^0)^2
                    \right)
                \right]
        =
    \\
    & =
        \dfrac{e \, b^{\alpha} \, |b^{5}|}{(2 \pi)^4 \sqrt{s^2 - 1}}
        \dfrac{\pi^2}{2i}
        (-1)
        \int_{0}^{+\infty} du \,
            \left\{
                \exp
                    \left[
                        \dfrac{i u}{4}
                        \left(
                            (y^1)^2 + (y^2)^2 - (y^3)^2 - (y^0)^2
                        \right)
                    \right]
            \right.
            -
        \\
    & \qquad \qquad \qquad \qquad \qquad \qquad \qquad \qquad
            -
            \left.
                \exp
                    \left[
                        -
                        \dfrac{iu}{4}
                        \left(
                            (y^1)^2 + (y^2)^2 - (y^3)^2 - (y^0)^2
                        \right)
                    \right]
            \right\}
\end{align*}
Where we have put $u = 1/q$. Here, the singularity at $q=0$ is controlled by the
oscillation in the exponent (although one can find the same result by other methods).

Using
\begin{align*}
    \int_{0}^{+\infty}
        \exp
            \left[
                i a x
            \right]
        dx
        & =
        P
        \left[
            \dfrac{i}{a}
        \right]
        +
        \pi \delta(a)
\end{align*}
we find the fields to be:
\begin{align*}
    a^{\alpha}(x,\tau)
    & =
        -
        \dfrac{e \, b^{\alpha} }{(2 \pi)^4 |b^{5}| \sqrt{s^2 - 1}}
        \dfrac{1}{2}
        \dfrac{8 \pi^2}
              {\left[ (y^1)^2 + (y^2)^2 - (y^3)^2 - (y^0)^2 \right]}
        =
    \\
    & =
        -
        \dfrac{e \, b^{\alpha}}{4 \pi^2 |b^{5}| \sqrt{s^2 - 1}}
        \dfrac{1}
              {
                \left[
                    (y^1)^2 + (y^2)^2 - (y^3)^2 - (y^0)^2
                \right]
              }
    \\
    \intertext{We have}
    y^{\mu} & =
        \left[
            x^0 ;
            x^1 , x^2 , \dfrac{x^3 - s\tau}{\sqrt{s^2 - 1}}
        \right]
    \\
    (y^1)^2 & + (y^2)^2 - (y^3)^2 - (y^0)^2 =
        (x^1)^2 + (x^2)^2 -
        \dfrac{(x^3 - s\tau)^2}{s^2 - 1} - (x^0)^2
        =
    \\
    & =
        x^2 - \dfrac{(b' \cdot x)^2}{b'^2}
        -
        \dfrac{(b \cdot x - b^2 \tau)^2}{(b'^2 - 1) b'^2}
\end{align*}
Thus we obtain the final form
\begin{align}
    a^{\alpha}(x,\tau) & =
        -
        \dfrac{e \, b^{\alpha}}{4 \pi^2 |b^{5}| \sqrt{b'^2 - 1}}
        \dfrac{1}
              {
                \left[
                        x^2
                    -
                        \dfrac{(b \cdot x)^2}{b^2}
                    -
                        \dfrac{(b \cdot x - b^2 \tau)^2}{b^2 (b^2 - 1)}
                \right]
              }
        \\
    \intertext{The corresponding 5D covariant form is then}
    a^{\alpha}(x,\tau)
    & =
        \dfrac{e n^{\alpha}}{4 \pi^2}
        \dfrac{1}
              {
                \left[
                        (n_{\alpha} x^{\alpha})^2
                    -
                        x_{\alpha} x^{\alpha}
                \right]
              }
\end{align}
where $n^{\alpha} = \dfrac{b^{\alpha}}{\sqrt{b_{\beta} b^{\beta}}}$

\subsection{Summary of fields generated by a Uniformly Moving Source}
We present a short summary of the results obtained in this section:

\begin{align}
    a^{\alpha}(x,\tau)
    & =
        \begin{cases}
                \dfrac{e n^{\alpha}}{4 \pi}
                \delta
                \left[
                    x_{\beta} x^{\beta} - (n_{\beta} x^{\beta})^2
                \right]
                \qquad
                \qquad
                &
                \text{Undershell $\sigma_5 = +1, n_{\alpha} n^{\alpha} = +1$}
            \\
            \\
                \dfrac{e n^{\alpha}}{4 \pi^2}
                \dfrac{1}
                    {
                        \left[
                            x_{\beta} x^{\beta} + (n_{\beta} x^{\beta})^2
                        \right]
                    }
                &
                \text{Supershell $\sigma_5 = +1, n_{\alpha} n^{\alpha} = -1$}
            \\
            \\
                \dfrac{e n^{\alpha}}{4 \pi}
                \delta
                \left[
                    x_{\beta} x^{\beta} + (n_{\beta} x^{\beta})^2
                \right]
                &
                \text{Under-spacelike $\sigma_5 = -1, n_{\alpha} n^{\alpha} = -1$}
            \\
            \\
                \dfrac{e n^{\alpha}}{4 \pi^2}
                \dfrac{1}
                    {
                        \left[
                            (n_{\beta} x^{\beta})^2
                            -
                            x_{\beta} x^{\beta}
                        \right]
                    }
                &
                \text{Super-spacelike $\sigma_5 = -1, n_{\alpha} n^{\alpha} = +1$}
        \end{cases}
\end{align}

In a a more general compact representation, we have
\begin{align}
    \label{eq:fields_solution_1}
    a^{\alpha}(x,\tau)
    & =
        \begin{cases}
                \dfrac{e n^{\alpha}}{4 \pi}
                \delta \left[ (n_{\beta} x^{\beta})^2 - \sigma_{5} x_{\beta} x^{\beta} \right]
            \qquad \qquad
            &
                \zeta = +1
            \\
            \\
                \dfrac{e n^{\alpha}}{4 \pi^2}
                \dfrac{1}
                      {
                        \left[
                            (n_{\beta} x^{\beta})^2 + \sigma_{5} x_{\beta} x^{\beta}
                        \right]
                      }
            \qquad \qquad
            &
                \zeta = -1
        \end{cases}
\end{align}
where:
\begin{align*}
    \zeta & = \sigma_{5} \cdot \varepsilon[b_{\alpha} b^{\alpha}]  =
              \sigma_{5} \cdot \varepsilon[b^2 + \sigma_{5} (b^{5})^2] \\
    n^{\alpha} & = \dfrac{b^{\alpha}}{|b_{\beta} b^{\beta}|}
\end{align*}

The various values for $\sigma_{5}$ and $\zeta$ are given in table
\ref{tab:RegionsOfSourceVelocitySummary}.

\begin{table}[htbp]
    \centering
        \begin{tabular}{|l|l|l|l|l|l|}
            \hline
            Metric  &   $\sigma_{5}$    &   Velocity region     &   Mass shell          &
                                                $\varepsilon(b_{\alpha} b^{\alpha})$    &   $\zeta$     \\
            \hline
            \hline
            $(4,1)$ &   $+1$            &   Undershell          &   $m^2 < M^2$         &   $+1$                                    &   $+1$        \\
            \hline
            $(4,1)$ &   $+1$            &   Supershell          &   $m^2 > M^2$         &   $-1$                                    &   $-1$        \\
            \hline
            $(3,2)$ &   $-1$            &   Under-spacelike     &   $-m^2 < M^2$        &   $-1$                                    &   $+1$        \\
            \hline
            $(3,2)$ &   $-1$            &   Super-spacelike     &   $-m^2 > M^2$        &   $+1$                                    &   $-1$        \\
            \hline
        \end{tabular}
    \caption{Regions of source velocity summary.}
    \label{tab:RegionsOfSourceVelocitySummary}
\end{table}

\subsection{Concatenation}
As we have pointed out above, the pre-Maxwell theory can be contracted to Maxwell
form by integration (as in \eqref{eq:maxwell_from_pre_maxwell_potential}) over $\tau$
(called \emph{concatenation})
Applying this procedure to the 5D pre-Maxwell fields which we have obtained above,
we find
\begin{align}
    \label{eq:concatenated_maxwell_fields}
    A^{\mu}(x)
    & =
        \int_{-\infty}^{+\infty}
            a^{\mu}(x,\tau)
            d\tau
    =
        \dfrac{e n^{\mu}}{4\pi}
        \dfrac{\theta[(n \cdot x)^2 - n^2 x^2]}
              {\sqrt{(n \cdot x)^2 - n^2 x^2}}
\end{align}
where
\begin{align*}
    n^2 & = n_{\mu} n^{\mu} = n_{\alpha} n^{\alpha} - \sigma_{5} (n^{5})^2 \\
\end{align*}
and $n_{\alpha} n^{\alpha} = \pm 1$ according to the velocity
regions of source motion (see table
\ref{tab:RegionsOfSourceVelocitySummary}). It should be emphasized
that \eqref{eq:concatenated_maxwell_fields} is a \emph{general}
Maxwell field, for all values of $n^2$. The solutions can be put into a more
specific form for the $3$ regions of $(3,1)$ spacetime:
\begin{align}
    A^{\mu}(x)
    & =
        \begin{cases}
                \dfrac{e n'^{\mu}}{4 \pi \sqrt{(n' \cdot x)^2 + x^2}}
                \qquad
                \qquad
                &
                n'^2 = -1
            \\
            \\
                \dfrac{e n'^{\mu}}{4 \pi}
                \dfrac{\theta[(n' \cdot x)^2 - x^2]}
                      {\sqrt{(n' \cdot x)^2 - x^2}}
                \qquad
                \qquad
                &
                n'^2 = +1
            \\
            \\
                \dfrac{e n'^{\mu}}{4 \pi |n' \cdot x|}
                \qquad
                \qquad
                &
                n'^2 = 0
        \end{cases}
\end{align}
where
\begin{align*}
    n'^{\mu} & = \dfrac{n^{\mu}}{|n_{\nu} n^{\nu}|}
\end{align*}

We will discuss the possibility of integrating on a smaller interval of $\tau$
(Land regularization \cite{MCLand1997}) adequate in some cases to reproduce the results
of ordinary Maxwell scattering.
%%
%%
%%
%%
%% END OF FIELDS
%%
%%
%%
%%
%%
%%

\section{Green functions}
%%%    \input{UMP_Green}
% =============================================================== %
% --------------------------------------------------------------- %
% -------------------> Green Functions <------------------------- %
% --------------------------------------------------------------- %
% =============================================================== %
\label{sec:green_functions}
In this section, Green functions for both $(4,1)$ and $(3,2)$
wave equations are given. Green functions for equations of
this type have been discussed \cite{Kythe1996,Gelfand1964_1,Polyanin2001}.
In particular, two distinct versions of the fundamental solution
for $(4,1)$ wave-equation have been given:
\begin{align}
    \label{eq:Maxwell_4_1_Green_function_no_theta}
    G_{4,1}(\vb{x},t)
    & =
    -
    \dfrac{1}{4 \pi^{2}}
    \dfrac{\theta(t - |\vb{x}|)}
          {[t^2 - \vb{x}^2]^{3/2}}
            \qquad \qquad
            &
            \text{cf. \cite{Polyanin2001}, based on \cite{Vladimirov1988}}
    \\
    \nonumber
    \\
    \label{eq:Maxwell_4_1_Green_function_with_theta}
    H_{4,1}(\vb{x},t)
    & =
    \dfrac{1}{2 \pi^2}
    \dfrac{d}{d t^2}
    \dfrac{\theta(t - |\vb{x}|)}
          {\sqrt{t^2 - \vb{x}^2}}
            \qquad \qquad
            &
            \text{cf. \cite{Kythe1996}}
\end{align}
The difference expression $(H_{4,1} - G_{4,1})(\vb{x},t)$ is a distribution
\begin{align*}
    (H_{4,1} - G_{4,1})(\vb{x},t)
    & =
        \dfrac{\delta(t - |\vb{x}|)}
              {2t \sqrt{t^2 - \vb{x}^2}}
\end{align*}

In the analysis below, on the other hand, we shall provide a distinct
derivation of the GF's for both $(4,1)$ and $(3,2)$ which are similar
to $H_{4,1}(x,t)$ as follows:
\begin{align*}
    g_{\sigma_{5}}(x,\tau)
    & =
        \lim \limits_{\epsilon \rightarrow 0^{+}}
        \dfrac{\sigma_{5}}{4 \pi^2}
        \dfrac{\partial}{\partial \epsilon}
        \dfrac{\theta [ - \sigma_{5} ( x^2 + \sigma_{5}\tau^2) + \epsilon]}
              {\sqrt{- \sigma_{5}( x^2 + \sigma_{5}\tau^2) + \epsilon}}
\end{align*}

$g_{\sigma_{5}}(x,\tau)$ contains a singular distribution term as well:
\begin{align*}
    \Delta_{\sigma_{5}}
    & =
        -
        \lim \limits_{\epsilon \rightarrow 0^{+}}
        \dfrac{\sigma_{5}}{4 \pi^2}
        \dfrac{\delta [ - \sigma_{5} ( x^2 + \sigma_{5}\tau^2) + \epsilon]}
              {\sqrt{- \sigma_{5}( x^2 + \sigma_{5}\tau^2) + \epsilon}}
\end{align*}

In the following sections, GF's are derived for the $(4,1)$ and $(3,2)$
wave equations, which are \emph{symmetric} in $t$. Then, we
apply the GF's to the current of a uniformly moving point source, and
re-derive the results of section \ref{sec:solutions}.
We shall describe the importance of the form of $\Delta_{\sigma_{5}}$
in the derivation of the fields.

% =============================================================== %
%                                                                 %
%                                                                 %
% ---------------> O(4,1) Green Function <----------------------- %
%                                                                 %
%                                                                 %
% =============================================================== %
\subsection{$(4,1)$ Green function}
With $\sigma_{5} = +1$ in \eqref{eq:hps_green_function_Fourier_solution},
we haves
\begin{align}
    \label{eq:4_1_green_function_fourier_rep}
    g_{+}(x,\tau)
    & =
        \dfrac{1}{(2\pi)^5}
        \int d^4k \, dk_{5} \,
            \dfrac{1}{k^2 + k_{5}^2}
            e^{i [k \cdot x + k_{5} \tau]}
    =
    \\
    & =
        \dfrac{1}{(2\pi)^5}
        \int d^3 \vb{k} \, dk_{5} \,
            e^{i[\vb{k} \cdot  \vb{x} + k_{5} \tau]}
            \int_{-\infty}^{+\infty}
            \dfrac{dk_{0}}{\vb{k}^2 + k_{5}^2 - k_{0}^2}
            e^{- i k_{0} t}
    \nonumber
\end{align}
The \emph{Cauchy Principal Part} of the $k_{0}$ integral is
\begin{align*}
    P
    \int_{-\infty}^{+\infty}
        \dfrac{dk_{0}}{\vb{k}^2 + k_{5}^2 - k_{0}^2}
        e^{- i k_{0} t}
    & =
        \pi i
        \varepsilon(-t)
        \dfrac{1}{2 \sqrt{\vb{k}^2 + k_{5}^2}}
        \left[
            +
            e^{+ i t \sqrt{\vb{k}^2 + k_{5}^2} }
            -
            e^{- i t \sqrt{\vb{k}^2 + k_{5}^2} }
        \right]
    =
    \\
    & =
        \varepsilon(t)
        \dfrac{\pi \sin \left( t \sqrt{\vb{k}^2 + k_{5}^2} \right) }
              {\sqrt{\vb{k}^2 + k_{5}^2}}
\end{align*}
We then have
\begin{align*}
    g_{+}(x,\tau)
    & =
        \dfrac{\pi \varepsilon(t)}{(2\pi)^5}
        \int d^3 \vb{k} \, dk_{5} \,
            \dfrac{\sin \left(  t \sqrt{\vb{k}^2 + k_{5}^2} \right) }
                  {\sqrt{\vb{k}^2 + k_{5}^2}}
            e^{i[\vb{k} \cdot  \vb{x} + k_{5} \tau]}
\end{align*}

We now orient the $\vb{k}, k_{5}$
space such the 4D "observation" vector $(\vb{x},\tau)$ is along
$k_{3}$ (one observes at time $\tau$ on the laboratory clock at the point $\vb{x}$).
Defining $l = \sqrt{\vb{k}^2 + k_{5}^2}$ and
$R = \sqrt{\vb{x}^2 + \tau^2}$, and using $\alpha, \theta$ and $\phi$ as
the 4D polar angles we find:
\begin{align*}
    k_{5} & = l \cos \alpha \\
    k_{3} & = l \sin \alpha \cos \theta \\
    \vb{k} \cdot \vb{x} + k_{5} \tau
    & =
              R k_{3} = R l \sin \alpha \cos \theta \\
    d^3 \vb{k} \, dk_{5}
    & =
              l^3 \sin^2 \alpha \sin \theta \, dl \, d\alpha \, d\theta \, d\phi
\end{align*}
In terms of these variables,
\begin{align*}
    g_{+}(x,\tau)
    & =
        \dfrac{\pi \varepsilon(t)}{(2\pi)^5}
        \int_{0}^{+\infty} l^3 \,  dl
        \int_{0}^{\pi}     \sin^2 \alpha \, d\alpha
        \int_{0}^{\pi}     \sin \theta   \, d\theta
        \int_{0}^{2\pi}    d\phi \,
            \dfrac{\sin (t l)}
                  {l}
            e^{i l R \sin \alpha \cos \theta}
    =
    \\
    & =
        \dfrac{\varepsilon(t) 4 \pi^2}{R (2\pi)^5}
        \int_{0}^{+\infty} l^3 \,  dl
        \int_{0}^{\pi}     \sin^2 \alpha \, d\alpha
            \dfrac{\sin(l R \sin \alpha)}{l \sin \alpha}
            \dfrac{\sin (t l)}
                  {l}
    =
    \\
    & =
        \dfrac{\varepsilon(t) 4 \pi^2}{R (2\pi)^5}
        \int_{0}^{+\infty} dl
        \int_{0}^{\pi}     d\alpha
            \sin(lt)
            \sin(lR \sin \alpha)
            l \sin \alpha
    =
    \\
    & =
        -
        \dfrac{\varepsilon(t) 4 \pi^2}{R (2\pi)^5}
        \dfrac{1}{R}
        \dfrac{\partial}{\partial R}
        \int_{0}^{+\infty} dl
        \int_{0}^{\pi}     d\alpha
            \sin(lt)
            \cos(lR \sin \alpha)
    =
    \\
\end{align*}
The choice of orientation in $k$-space resulted in a first order
derivative with respect to the "4D observation point" $R$. This form
is important in convergence of the UMS solution. The $\alpha$
integral is simply $\pi J_{0}(lR)$ and using the \emph{sine
transform} of $J_{0}(x)$ (cf. \cite{Arfken1985})
\begin{align}
    \label{eq:J_0_sine_transform}
    J_{0}(x)
    & =
        \dfrac{2}{\pi}
        \int_{1}^{+\infty}
            \dfrac{\sin(x s)}{\sqrt{s^2 - 1}}
            ds
\end{align}
we find:
\begin{align*}
    g_{+}(x,\tau)
    & =
        -
        \dfrac{8 \pi^2\varepsilon(t)}{(2\pi)^5}
        \dfrac{1}{R}
        \dfrac{\partial}{\partial R}
        \int_{0}^{+\infty} dl
        \int_{1}^{+\infty} ds \,
            \sin(lt)
            \dfrac{ \sin(l R s)}{\sqrt{s^2 - 1}}
\end{align*}
Changing the order of integration and noting that the $l$ integrand is symmetric
under $l \rightarrow -l$ we have
\begin{align*}
    g_{+}(x,\tau)
    & =
        -
        \dfrac{8 \pi^2\varepsilon(t)}{(2\pi)^5}
        \dfrac{1}{R}
        \dfrac{\partial}{\partial R}
        \int_{1}^{+\infty}
            \dfrac{ds}{\sqrt{s^2 - 1}}
        \dfrac{1}{2}
        \int_{-\infty}^{+\infty} dl
            \sin(lt)
            \sin(l R s)
    =
    \\
    & =
        -
        \dfrac{4 \pi^2\varepsilon(t)}{(2\pi)^5}
        \dfrac{1}{R}
        \dfrac{\partial}{\partial R}
        \int_{1}^{+\infty}
            \dfrac{ds}{\sqrt{s^2 - 1}}
            \dfrac{2\pi}{(-4)}
            \left[
                2 \delta(t + Rs)
                -
                2 \delta(t - Rs)
            \right]
\end{align*}
Since $Rs > 0$, then using
\begin{align*}
    \delta(t + Rs) - \delta(t - Rs)
    & =
        -
        \varepsilon(t)
        \left[
            \delta(t + Rs) + \delta(t - Rs)
        \right]
\end{align*}
we find:
\begin{align*}
    g_{+}(x,\tau)
    & =
        \dfrac{4 \pi^3\varepsilon^2(t)}{(2\pi)^5}
        \dfrac{1}{R}
        \dfrac{\partial}{\partial R}
        \int_{1}^{+\infty}
            \dfrac{ds}{\sqrt{s^2 - 1}}
            \left[
                \delta(t + Rs)
                +
                \delta(t - Rs)
            \right]
    =
    \\
    & =
        \dfrac{4 \pi^3}{(2\pi)^5}
        \dfrac{1}{R}
        \dfrac{\partial}{\partial R}
            \dfrac{1}{R}
            \dfrac{1}{\sqrt{t^2/R^2 - 1}}
            \left[
                \theta(-t/R-1)
                +
                \theta(+t/R-1)
            \right]
    =
    \\
    & =
        \dfrac{1}{8\pi^2}
        \dfrac{1}{R}
        \dfrac{\partial}{\partial R}
        \dfrac{\theta(t^2 - R^2)}{\sqrt{t^2 - R^2}}
\end{align*}
We have $t^2 - R^2 = t^2 - \vb{x}^2 - \tau^2 = - x_{\alpha} x^{\alpha}$,
and since $R > 0$, we can use $1/R \partial/\partial R = 2 \partial/\partial R^2$,
which is \emph{linear} in $x_{\alpha} x^{\alpha}$, thus:
\begin{align}
    \label{eq:O_4_1_green_function}
    g_{+}(x,\tau)
    & =
        -
        \lim \limits_{\epsilon \rightarrow 0^{+}}
        \dfrac{1}{4\pi^2}
        \dfrac{\partial}{\partial \epsilon}
        \dfrac{\theta(- x_{\alpha} x^{\alpha} + \epsilon)}
              {\sqrt{- x_{\alpha} x^{\alpha} + \epsilon}}
\end{align}

% =============================================================== %
%                                                                 %
%                                                                 %
% ---------------> O(3,2) Green Function <----------------------- %
%                                                                 %
%                                                                 %
% =============================================================== %
\subsection{$(3,2)$ Green function}
We shall repeat the procedure for $\sigma_{5} = -1$, as follows:
\begin{align}
    \label{eq:3_2_green_function_fourier_rep}
    g_{-}(x,\tau)
    & =
        \dfrac{1}{(2\pi)^5}
        \int d^4k \, dk_{5} \,
            \dfrac{1}{k^2 - k_{5}^2}
            e^{i [k \cdot x - k_{5} \tau]}
    =
    \\
    & =
        \dfrac{1}{(2\pi)^5}
        \int_{\mathbb{R}^3} d^3 \vb{k} \,
            e^{i[\vb{k} \cdot  \vb{x}]}
            \int_{\mathbb{R}^2}
            \dfrac{dk_{5} \, dk_{0}}{\vb{k}^2 - k_{5}^2 - k_{0}^2}
            e^{- i [k_{0} t + k_{5} \tau]}
    \nonumber
\end{align}
The integration is separated into the two subspaces $\mathbb{R}^3$ for the
spatial coordinates, and $\mathbb{R}^2$ for the temporal coordinates.
We shall use polar coordinates in \emph{both} spaces, using the following
substitutions:
\begin{align*}
    k^2 & = k_{1}^2 + k_{2}^2 + k_{3}^2  \qquad &
                    d^3 \vb{k} = k^2 \sin \theta dk \, d\theta \, d\phi \\
    r^2 & = x^2 + y^2 + x^2              \qquad &
                    \vb{k} \cdot \vb{x} = k r \cos \theta               \\
    l^2 & = k_{0}^2 + k_{5}^2            \qquad &
                    dk_{0} \, dk_{5} = l \, dl \, d\alpha               \\
    s^2 & = t^2 + \tau^2                 \qquad &
                    k_{0}t + k_{5}\tau = s l \cos \alpha                \\
\end{align*}
The integral then takes the form
\begin{align*}
    g_{-}(x,\tau)
    & =
        \dfrac{1}{(2\pi)^5}
        \int_{0}^{+\infty} k^2          \,  dk
        \int_{0}^{\pi}     \sin \theta  \,  d\theta
        \int_{0}^{2\pi}                     d\phi   \,
            e^{i k r \cos \theta}
            \int_{0}^{+\infty}  l       \,  dl
            \int_{0}^{2\pi}                 d\alpha \,
            \dfrac{e^{i l s \cos \alpha}}{k^2 - l^2}
\end{align*}
We can integrate immediately on $\phi$, $\theta$ and $\alpha$ as follows:
\begin{align*}
    \int_{0}^{2\pi} d\phi   & = 2\pi                        , \qquad
    \int_{0}^{\pi}  d\theta   = \dfrac{2 \sin(kr)}{kr}      , \qquad
    \int_{0}^{2\pi} d\alpha   = 2\pi J_{0} (l s)
    \intertext{and thus:}
    g_{-}(x,\tau)
    & =
        \dfrac{1}{(2\pi)^5}
        \dfrac{8\pi^2}{r}
        \int_{0}^{+\infty} k \, dk \,
            \sin(kr)
            \int_{0}^{+\infty}  l \, dl \,
            \dfrac{J_{0}(ls)}{k^2 - l^2}
    =
    \\
    & =
        -
        \dfrac{1}{(2\pi)^3}
        \dfrac{2}{r}
        \dfrac{\partial}{\partial r}
        \int_{0}^{+\infty} l \, dl \,
            J_{0}(ls)
        \int_{0}^{+\infty}      dk \,
            \dfrac{\cos(kr)}{k^2 - l^2}
    =
    \\
    & =
        -
        \dfrac{1}{(2\pi)^3}
        \dfrac{1}{r}
        \dfrac{\partial}{\partial r}
        \int_{0}^{+\infty} l \, dl \,
            J_{0}(ls)
        \int_{-\infty}^{+\infty} dk \,
            \dfrac{\cos(kr)}{k^2 - l^2}
\end{align*}
The \emph{Principal Part} value of the $k$ integral is:
\begin{align*}
    P
    \int_{-\infty}^{+\infty}
        \dfrac{\cos(kr)}{k^2 - l^2}
        dk
    & =
        P.P.
        \,
        \Re
        \left[
            \int_{-\infty}^{+\infty}
                \dfrac{e^{ikr}}{k^2 - l^2}
                dk
        \right]
    =
    \\
    & =
        \Re
        \left\{
            \dfrac{i \pi}
                  {2 l}
            \left[
                e^{i l r}
                -
                e^{- i l r}
            \right]
        \right\}
    =
        -
        \dfrac{\pi}{l}
        \sin(lr)
\end{align*}
Thus, the expression for $g_{-}(x,\tau)$ one obtains the form
\begin{align*}
    g_{-}(x,\tau)
    & =
        \dfrac{\pi}{(2\pi)^3}
        \dfrac{1}{r}
        \dfrac{\partial}{\partial r}
        \int_{0}^{+\infty} l \, dl \,
            J_{0}(ls)
            \dfrac{\sin(lr)}{l}
    =
    \\
    & =
        \dfrac{\pi}{(2\pi)^3}
        \dfrac{1}{r}
        \dfrac{\partial}{\partial r}
        \int_{0}^{+\infty} dl \,
            J_{0}(ls) \sin(lr)
\end{align*}
Once again, using the sine transform of $J_{0}(x)$ (see \eqref{eq:J_0_sine_transform}),
we have
\begin{align*}
    g_{-}(x,\tau)
    & =
        \dfrac{\pi}{(2\pi)^3}
        \dfrac{2}{\pi}
        \dfrac{1}{r}
        \dfrac{\partial}{\partial r}
        \int_{1}^{+\infty}
            \dfrac{du}{\sqrt{u^2 - 1}}
        \dfrac{1}{2}
        \int_{-\infty}^{+\infty} dl \,
            \sin(ls u) \sin(lr)
    =
    \\
    & =
        \dfrac{1}{(2\pi)^3}
        \dfrac{1}{r}
        \dfrac{\partial}{\partial r}
        \int_{1}^{+\infty}
            \dfrac{du}{\sqrt{u^2 - 1}}
            \dfrac{2 \pi}{(-4)}
            \left[
                2 \delta(su + r)
                -
                2 \delta(su - r)
            \right]
    =
    \\
    & =
        -
        \dfrac{\pi}{(2\pi)^3}
        \dfrac{1}{r}
        \dfrac{\partial}{\partial r}
        \int_{1}^{+\infty}
            \dfrac{du}{\sqrt{u^2 - 1}}
            \epsilon(-s)
            \left[
                \delta(su + r)
                +
                \delta(su - r)
            \right]
\end{align*}
However, $su \geq 0$ and $r \geq 0$, and thus, the first term $\delta(su+r)$ cancels
identically, leaving:
\begin{align*}
    g_{-}(x,\tau)
    & =
        \dfrac{1}{2(2\pi)^2}
        \dfrac{1}{r}
        \dfrac{\partial}{\partial r}
        \dfrac{1}{s}
        \dfrac{\theta(r/s - 1)}
              {\sqrt{r^2/s^2 - 1}}
    =
    \\
    & =
        \dfrac{1}{2(2\pi)^2}
        \dfrac{1}{r}
        \dfrac{\partial}{\partial r}
        \dfrac{\theta(r - s)}
              {\sqrt{r^2 - s^2}}
\end{align*}
Using the same arguments that were made for the $(4,1)$ case, we have:
\begin{align}
    \label{eq:O_3_2_green_function}
    g_{-}(x,\tau)
    & =
        \lim \limits_{\epsilon \rightarrow 0^{+}}
        \dfrac{1}{4\pi^2}
        \dfrac{\partial}{\partial \epsilon}
        \dfrac{\theta(x_{\alpha} x^{\alpha} + \epsilon)}
              {\sqrt{x_{\alpha} x^{\alpha} + \epsilon}}
\end{align}
where $x_{\alpha} x^{\alpha} = x^2 - \tau^2$ in this case.

Combining both $(4,1)$ and $(3,2)$ cases, we obtain:
\begin{align}
    \label{eq:green_functions_final}
    g_{\sigma_{5}}(x,\tau)
    & =
        \lim \limits_{\epsilon \rightarrow 0^{+}}
        \dfrac{\sigma_{5}}{4\pi^2}
        \dfrac{\partial}{\partial \epsilon}
        \dfrac{\theta[- \sigma_{5} (x^2 + \sigma_{5} \tau^2) + \epsilon]}
              {\sqrt{- \sigma_{5} (x^2 + \sigma_{5} \tau^2) + \epsilon}}
    =
        \lim \limits_{\epsilon \rightarrow 0^{+}}
        \dfrac{\sigma_{5}}{4\pi^2}
        \dfrac{\partial}{\partial \epsilon}
        \dfrac{\theta[- \sigma_{5} x_{\alpha} x^{\alpha}  + \epsilon]}
              {\sqrt{ - \sigma_{5} x_{\alpha} x^{\alpha}  + \epsilon}}
\end{align}

The factor 2 between \eqref{eq:green_functions_final} and
\eqref{eq:Maxwell_4_1_Green_function_with_theta} stems from the fact
that \emph{both} retarded and advanced $t$ are used, picking up
additional contribution from the future of $t$. An attempt to
provide GF's which are $\tau$ retarded in $(4,1)$ and $(3,2)$
is currently under study.

% =============================================================== %
%                                                                 %
%                                                                 %
% ---------------> UMS solutions with GF <----------------------- %
%                                                                 %
%                                                                 %
% =============================================================== %
\subsection{Fields solution through GF's}
We shall now apply the GF's \eqref{eq:green_functions_final} to the
current generated by a uniformly moving point source. Recalling the
UMS path \eqref{eq:ums_worldline}, generating the current
\eqref{eq:point_source_current}, we shall use
\eqref{eq:hps_potential_solution_using_green_function} to find the
fields:
\begin{align}
    \label{eq:field_integral_through_green_function}
    a_{\sigma_{5}}^{\alpha}(x,\tau)
    & =
        \lim \limits_{\epsilon \rightarrow 0^{+}}
        \dfrac{e \sigma_{5}}{4\pi^2}
        \dfrac{\partial}{\partial \epsilon}
        \int_{-\infty}^{+\infty} d\tau' \,
        \,
        b^{\alpha}
        \,
        \dfrac{\theta[- \sigma_{5} ((x - b\tau') + \sigma_{5} (\tau - b^{5} \tau')^2) + \epsilon]}
              {\sqrt{- \sigma_{5} ((x - b\tau') + \sigma_{5} (\tau - b^{5} \tau')^2) + \epsilon}}
\end{align}
Clearly, the limits of integration depend on the coefficients of the
quadratic argument:
\begin{align*}
    p(\tau')
    & =
        - \sigma_{5} ((x - b\tau') + \sigma_{5} (\tau - b^{5} \tau')^2) + \epsilon
    =
    \\
    & =
        - \sigma_{5}
        \left[
            (b^2 + \sigma_{5} (b^{5})^2) \tau'^2
            -
            2 (b \cdot x + \sigma_{5} b^{5} \tau) \tau'
            +
            x^2 + \sigma_{5} \tau^2
        \right]
        +
        \epsilon
    =
    \\
    & =
            - \sigma_{5} b_{\alpha} b^{\alpha} \tau'^2
        +
            2 \sigma_{5} (b_{\alpha} x^{\alpha}) \tau'
        -
            \sigma_{5}
            \left[
                x_{\alpha} x^{\alpha}
                -
                \sigma_{5}
                \epsilon
            \right]
\end{align*}
Thus, the polarity of the quadratic form $p(\tau)$ depends on the
sign of $\zeta \equiv \sigma_{5}  \varepsilon(b_{\alpha}
b^{\alpha})$, which we shall treat individually.

% =============================================================== %
%                                                                 %
%                                                                 %
% ---------------> UMS solutions with GF <----------------------- %
%                                                                 %
%                                                                 %
% =============================================================== %
\subsection{UMS fields for $\zeta = +1$}
We have $- \sigma_{5} b_{\alpha} b^{\alpha} < 0$. The condition
$p(\tau') > 0$ is then limited to $\tau_{1} < \tau < \tau_2$ where
$\tau_{1,2}$ are the \emph{roots} of $p(\tau')$:
\begin{align}
    \label{eq:p_tau_roots}
    \tau'_{1,2}
    & =
        \dfrac{- b_{\alpha} x^{\alpha}
               \pm
               \sqrt{(b_{\alpha} x^{\alpha})^2
                     -
                     b_{\alpha} b^{\alpha}
                     (x_{\alpha} x^{\alpha} - \sigma_{5} \epsilon)
                    }
              }
              {-b_{\alpha} b^{\alpha}}
\end{align}
the fields become
\begin{align}
    \label{eq:a_field_integral_closed}
    a_{\sigma_{5}}^{\alpha}(x,\tau)
    & =
        \lim \limits_{\epsilon \rightarrow 0^{+}}
        \dfrac{e \sigma_{5} b^{\alpha}}{4\pi^2}
        \dfrac{\partial}{\partial \epsilon}
        \int_{\tau_{1}}^{\tau_{2}} d\tau' \,
        \,
        \dfrac{\theta[p(\tau')]}
              {\sqrt{p(\tau')}}
\end{align}

Clearly, the integral would be become zero \emph{identically} if
$\tau_{1,2}$ are \emph{complex}, thus:
\begin{align*}
    (\tau'_{1} - \tau'_{2})^2
    & =
        \dfrac{4}{(b_{\alpha} b^{\alpha})^2}
        \left[
            (b_{\alpha} x^{\alpha})^2
            -
            b_{\alpha} b^{\alpha}
            (x_{\alpha} x^{\alpha} - \sigma_{5} \epsilon)
        \right]
    > 0
\end{align*}

We can now write $p(\tau')$ as follows:
\begin{align}
    p(\tau')
    \label{eq:p_tau_with_R2_1}
    & \equiv
        \dfrac{\sigma_{5}}{b_{\alpha} b^{\alpha}}
        R^2
        -
        A^2(\tau' - B)^2
    \\
    \intertext{where}
    R^2
    & =
        (b_{\alpha} x^{\alpha})^2
        -
        b_{\alpha} b^{\alpha}
        \left( x_{\alpha} x^{\alpha} - \sigma_{5} \epsilon \right)
    \qquad
    A = \sqrt{\sigma_{5} b_{\alpha} b^{\alpha}}
    \qquad
    B = \dfrac{b_{\alpha} x^{\alpha}}{b_{\alpha} b^{\alpha}}
\end{align}
where $R^2 > 0$ is a requirement for the integral to be non-zero,
and $A^2 > 0$ in this case. After making the substitution
\begin{align*}
    \sqrt{\dfrac{\sigma_{5}}{b_{\alpha} b^{\alpha}}} R \tanh \beta & = A (\tau' - B)
\end{align*}
we have
\begin{align*}
    a_{\sigma_{5}}^{\alpha}(x,\tau)
    & =
        \lim \limits_{\epsilon \rightarrow 0^{+}}
        \sqrt{\dfrac{e \sigma_{5}}{b_{\alpha} b^{\alpha}}}
        \dfrac{\sigma_{5} b^{\alpha}}{4\pi^2}
        \dfrac{\partial}{\partial \epsilon}
        \theta(R^2)
        \int_{-\infty}^{+\infty}
            \dfrac{R \, d\beta}{A \cosh^2 \beta}
        \,
        \dfrac{1}
              {\sqrt{ \left(\sigma_{5}/b_{\alpha} b^{\alpha} \right) R^2
                      \left[1 - \tanh^2 \beta \right]
                    }
              }
    =
    \nonumber
    \\
    & =
        \lim \limits_{\epsilon \rightarrow 0^{+}}
        \dfrac{1}{\sqrt{\sigma_{5} b_{\alpha} b^{\alpha}}}
        \dfrac{e \sigma_{5} b^{\alpha}}{4\pi^2}
        \dfrac{\partial}{\partial \epsilon}
        \theta(R^2)
        \int_{-\infty}^{+\infty}
            \dfrac{d\beta}{\cosh \beta}
    \\
\end{align*}
The remaining $\beta$ integral is a constant and equal to $\pi$ (easily
verified by substituting $u = e^{\beta}$). Thus
\begin{align*}
    a_{\sigma_{5}}^{\alpha}(x,\tau)
    & =
        \lim \limits_{\epsilon \rightarrow 0^{+}}
        \dfrac{e \sigma_{5} b^{\alpha}}
              {4\pi \sqrt{b_{\alpha} b^{\alpha}}}
        \dfrac{\partial}{\partial \epsilon}
        \theta(R^2)
\end{align*}
where $\partial_{\epsilon} \theta(R^2) = \sigma_{5} b_{\alpha}
b^{\alpha} \delta(R^2)$, which gives the final form for
$a_{\sigma_{5}}(x,\tau)$ for this case:
\begin{align}
    a^{\alpha}_{\sigma_{5}}(x,\tau)
    & =
        \dfrac{e b^{\alpha} (\sigma_{5} b_{\beta} b^{\beta})}
              {4 \pi \sqrt{\sigma_{5} b_{\alpha} b^{\alpha}}}
        \delta
        \left[
            (b_{\alpha} x^{\alpha})^2
            -
            b_{\alpha} b^{\alpha} x_{\alpha} x^{\alpha}
        \right]
\end{align}
where the limit of $\epsilon \rightarrow 0^{+}$ was taken
explicitly. Defining the normalized 5D velocity $n^{\alpha} =
b^{\alpha} / \sqrt{\sigma_{5} b_{\beta} b^{\beta}}$, we obtain the
solution consistent with $\zeta = 1$ of \eqref{eq:fields_solution_1}
\begin{align}
    a^{\alpha}_{\sigma_{5}}(x,\tau)
    & =
        \dfrac{e n^{\alpha}}
              {4 \pi}
        \delta
        \left[
            (n_{\alpha} x^{\alpha})^2
            -
            \sigma_{5} x_{\alpha} x^{\alpha}
        \right]
\end{align}
where it is stressed again that $\sigma_{5}$ appears implicitly in
the scalar products such as $n_{\alpha} x^{\alpha}$.

% =============================================================== %
%                                                                 %
%                                                                 %
% ---------------> UMS solutions with GF <----------------------- %
%                                                                 %
%                                                                 %
% =============================================================== %
\subsection{UMS fields for $\zeta = +1$}
We shall repeat the analysis of the last section for the case of
$-\sigma_{5} b_{\alpha} b^{\alpha} > 0$. The roots
\eqref{eq:p_tau_roots} are applicable in the present case as well.
However, the range of integration of
\eqref{eq:field_integral_through_green_function} now reveals that
$p(\tau') > 0$ for the \emph{exterior} region $\tau' \in (-\infty,
\tau'_{1}) \cup (\tau'_{2}, +\infty)$. Therefore,
\eqref{eq:p_tau_with_R2_1} now becomes:
\begin{align}
    \label{eq:p_tau_with_R2_2}
    p(\tau')
    & =
        \left(
            - \sigma_{5} b_{\alpha} b^{\alpha}
        \right)
        \left(
            \tau' - \dfrac{b_{\alpha} x^{\alpha}}{b_{\alpha} b^{\alpha}}
        \right)^2
        -
        \left(
            - \dfrac{\sigma_{5}}{b_{\alpha} b^{\alpha}}
        \right)
        \left[
            (b_{\alpha} x^{\alpha})^2
            -
            b_{\alpha} b^{\alpha}
            ( x_{\alpha} x^{\alpha} - \sigma_{5} \epsilon)
        \right]
    \nonumber
    =
    \\
    & =
        A^2(\tau' - B)^2
        -
        \left(
            - \dfrac{\sigma_{5}}{b_{\alpha} b^{\alpha}}
        \right)
        R^2
\end{align}
where in this case
\begin{align*}
    A & = \sqrt{- \sigma_{5} b_{\alpha} b^{\alpha}}
\end{align*}

However, in order that the field integral
\eqref{eq:a_field_integral_closed} converges, we shall define
$p_{\lambda}(\tau')$ as follows:
\begin{align}
    p_{\rho}(\tau') & =
        p(\tau') + \rho^2
\end{align}

The field integral \eqref{eq:a_field_integral_closed} obtains the
form
\begin{align}
    a_{\sigma_{5}}^{\alpha}(x,\tau)
    & =
        \dfrac{e \sigma_{5} b^{\alpha}}{4 \pi^2}
        \lim\limits_{
                        \substack{  \rho     \to 0^{+}
                                    \\
                                    \epsilon \to 0^{+}
                                 }
                    }
        \dfrac{\partial}{\partial \epsilon}
        \int_{-\infty}^{+\infty}
            \dfrac{\theta(p(\tau'))}
                  {\sqrt{p_{\rho}(\tau')}}
            d\tau'
    =
    \nonumber
    \\
    & =
        \label{eq:a_field_integral_open}
        \dfrac{e \sigma_{5} b^{\alpha}}{4 \pi^2}
        \lim\limits_{
                        \substack{  \rho     \to 0^{+}
                                    \\
                                    \epsilon \to 0^{+}
                                 }
                    }
        \int_{-\infty}^{+\infty}
        \left[
            \dfrac{\delta(p(\tau'))}
                  {\sqrt{p_{\rho}(\tau')}}
            -
            \dfrac{1}{2}
            \dfrac{\theta(p(\tau'))}
                  {[p_{\rho}(\tau')]^{3/2}}
        \right]
        \dfrac{\partial p(\tau')}{\partial \epsilon}
\end{align}
where we used that fact that $\partial_{\epsilon} p(\tau') =
\partial_{\epsilon} p_{\rho}(\tau')$.
The first $\delta(p(\tau'))$ term breaks up over the roots of
$p(\tau')$:
\begin{align}
    \label{eq:delta_function_term}
    \dfrac{\delta(p(\tau'))}
            {\sqrt{p_{\rho}(\tau')}}
    & =
        \dfrac{1}{A^2}
        \dfrac{\delta(\tau - \tau'_{1}) + \delta(\tau - \tau'_{2})}{|\tau'_{1} - \tau'_{2}|}
        \dfrac{1}{\sqrt{p_{\rho}(\tau')}}
\end{align}
We also have
\begin{align}
    \label{eq:epsilon_p_tau_derivative}
    \partial_{\epsilon}p(\tau')
    & =
        \partial_{\epsilon} p_{\rho}(\tau')
    =
        1
\end{align}

Integrating the first $\delta$-term, and combining
\eqref{eq:epsilon_p_tau_derivative} with
\eqref{eq:delta_function_term} and \eqref{eq:a_field_integral_open},
we obtain
\begin{align}
    \int_{-\infty}^{+\infty}
        \dfrac{\delta(p(\tau'))}
              {\sqrt{p_{\rho}(\tau')}}
              \dfrac{\partial p(\tau')}{\partial \epsilon}
        d\tau'
    & =
        \dfrac{1}{A^2}
        \dfrac{|b_{\alpha} b^{\alpha}|}{2 R}
        \left[
            \dfrac{1}{\sqrt{p_{\rho}(\tau')}}
            \Bigg|_{\tau' = \tau'_{1}}
            +
            \dfrac{1}{\sqrt{p_{\rho}(\tau')}}
            \Bigg|_{\tau' = \tau'_{2}}
        \right]
        \nonumber
    \\
    & =
        \label{eq:delta_function_integral}
        \dfrac{1}{R}
        \dfrac{1}{\rho}
\end{align}
which diverges as $1/\rho$. The second term can be integrated with
the substitution
\begin{align*}
    C
    \cosh \beta
    & =
        A(\tau' - R)
    \qquad
    \text{where}
    \qquad
    C
    =
        \left[
            \sqrt{
                \dfrac{- \sigma_{5}}{b_{\alpha} b^{\alpha}}
                R^2
                -
                \lambda^2
            }
        \,
        \right]
\end{align*}
Thus
\begin{align}
    \label{eq:theta_function_integral}
    \int_{-\infty}^{+\infty}
        \dfrac{\theta(p(\tau'))}
              {[p_{\rho}(\tau')]^{3/2}}
        d\tau'
    & =
        \dfrac{2 C}{A}
        \theta(R^2)
        \int_{\beta_{0}}^{+\infty}
        \sinh \beta \, d\beta \,
        \dfrac{1}{[C^2 (\cosh^2 \beta - 1)]^{3/2}}
    =
    \\
    & =
        2
        \dfrac{\theta(R^2)}{A C^2}
        \int_{\beta_{0}}^{+\infty}
        \dfrac{d\beta}{\sinh^2 \beta}
    =
        2
        \dfrac{\theta(R^2)}{A C^2}
        (-1)
        \coth \beta \Bigg|_{\beta_{0}}^{+\infty}
    =
    \nonumber
    \\
    & =
        2
        \dfrac{\theta(R^2)}{A C^2}
        \left[
            \coth \beta_{0} - 1
        \right]
    \nonumber
\end{align}
The $\beta_{0}$ lower bound is given by
\begin{align*}
    \sinh^2 \beta_{0} & = \dfrac{\rho^2}{C^2}
    \Longrightarrow
    \coth \beta_{0}
    =
        \sqrt{1 + \dfrac{1}{\sinh^2 \beta_{0}}}
    =
        \dfrac{1}{\rho}
        \sqrt{C^2 + \rho^2}
    =
        \dfrac{C}{\rho}
        +
        O(\rho)
\end{align*}
which provides the complete solution for the second term
\begin{align}
    \label{eq:theta_function_integral_2}
    \int_{-\infty}^{+\infty}
        \dfrac{\theta(p(\tau'))}
              {[p_{\rho}(\tau')]^{3/2}}
        d\tau'
    & =
        2
        \dfrac{\theta(R^2)}{A C^2}
        \left[
                \dfrac{C}{\rho}
            -
                1
        \right]
    =
    \\
    & =
        2 \theta(R^2)
        \left[
            \dfrac{1}{\sqrt{R^2 + \sigma_{5} \rho^2 b_{\alpha} b^{\alpha} }}
            \dfrac{1}{\rho}
            -
            \dfrac{1}
                  {
                    \sqrt{- \sigma_{5} b_{\alpha} b^{\alpha}}
                    (
                        - \sigma_{5} R^2 / b_{\alpha} b^{\alpha}
                    -
                        \rho^2
                    )
                  }
        \right]
\end{align}
The sum of the $\delta$-term and the smooth term becomes
\begin{align*}
    a_{\sigma_{5}}^{\alpha}(x,\tau)
    & =
        \dfrac{e \sigma_{5} b^{\alpha}}{4\pi^2}
        \lim\limits_{
                        \substack{  \rho     \to 0^{+}
                                    \\
                                    \epsilon \to 0^{+}
                                 }
                    }
        \theta(R^2)
        \left[
                \dfrac{1}{R} \dfrac{1}{\rho}
            -
            \left(
                \dfrac{1}{\sqrt{R^2 + \sigma_{5} \rho^2 b_{\alpha} b^{\alpha} }}
                \dfrac{1}{\rho}
                -
                \dfrac{1}
                    {
                        \sqrt{- \sigma_{5} b_{\alpha} b^{\alpha}}
                        (
                            - \sigma_{5} R^2 / b_{\alpha} b^{\alpha}
                        -
                            \rho^2
                        )
                    }
            \right)
        \right]
    \\
    & =
        \dfrac{e b^{\alpha}}
              {4\pi^2 \sqrt{- \sigma_{5} b_{\alpha} b^{\alpha}}}
        \dfrac{b_{\alpha} b^{\alpha}}
              {
                \left[
                    (b_{\alpha} x^{\alpha})^2
                    -
                    (b_{\alpha} b^{\alpha}) (x_{\alpha} x^{\alpha})
                \right]
              }
\end{align*}
Once again, defining $n^{\alpha} = b^{\alpha} / |b_{\alpha}
b^{\alpha}|$, we obtain the final result for $\zeta = -1$:
\begin{align}
    \label{eq:smooth_fields_through_green_functions}
    a^{\alpha}_{\sigma_{5}}(x,\tau)
    & =
        \dfrac{e n^{\alpha}}{4\pi^2}
        \dfrac{
                \theta(
                        (n_{\alpha} x^{\alpha})^2
                        +
                        \sigma_{5} x_{\alpha} x^{\alpha}
                      )
              }
              {
                \left[
                    (n_{\alpha} x^{\alpha})^2
                    +
                    \sigma_{5} x_{\alpha} x^{\alpha}
                \right]
              }
\end{align}
%%
%%
%%
%%
%%
%%
%%
%%
%% END OF GREEN-FUNCTIONS
%%
%%
%%
%%
%%
%%
%%
%%

\section{Conclusions}
%%%    \input{UMP_Conclusions}
%=================================================================%
%-----------------------------------------------------------------%
%- Project...........: UMP - Uniform Motion Paper.                %
%- File..............: UMP_Conclusions.tex                        %
%- Synopsis..........: Conclusions for the UMP paper.             %
%-                                                                %
%- Author(s).........: Larry Horwitz and Jigal Aharonovich.       %
%-----------------------------------------------------------------%
%=================================================================%

\label{sec:conclusions}

The $a$-fields (see \eqref{eq:fields_solution_1}) generated by a
uniformly moving point source in $(4,1)$ and $(3,2)$ offshell
electrodynamics clearly resemble the expected UMS fields in a 5D
Maxwell electrodynamics. However, the latter, generally in a
framework of relativistic dynamics, normally regarded as producing
fields from \emph{timelike} sources only. Stueckelberg based
offshell electrodynamics, on the other hand, that are based on a 4D
dynamics parameterized along an invariant parameter, have no
apparent limit on the region of the source velocity, though
normally, the equations are set with constant $z^{5}(\tau) \equiv
\tau \Rightarrow b^{5} \equiv 1$. Moreover, the 4D dynamics place no
\emph{a priori} restriction on $b'^{\mu} \equiv \frac{b^{\mu}}{b^{5}}$, and
all cases of $b'^2 < 0$ or $b'^2 > 0$ were shown. However, the two
forms of the fields given in \eqref{eq:fields_solution_1} differ
dramatically, and in particular, one is required to explain the
$\delta$-functions fields found for the 5D spacelike and timelike
regions of source velocity for the $(4,1)$ and $(3,2)$ flat metric
equations, respectively.

The $\delta$-function fields ($\zeta = -1$) have support on a 4D
null surface given by $x_{\alpha} x^{\alpha} - \sigma_{5}
(n_{\alpha} x^{\alpha})^2 = 0$, which is orthogonal to the direction
of motion of the source $n^{\alpha}$. Thus, this null surface is
actually the $(3,1)$ light-cone, as can easily be observed in the
frame $n^{\alpha} = [0; 0,0,0, 1]$, in which case, both $(4,1)$ and
$(3,2)$ fields reduce to the Maxwell time-symmetric GF
$\delta(x^2)$. This reflects the choice of Principal Part which was
taken in the derivation of those fields. These singular fields are
in fact the analog for a 4D UMS field of a spacelike moving source.

In a subsequent study, we plan to show that when a Lorentz force derived
from those fields is applied to a test particle, it produces a finite
force in an infinitesimally short $\tau$ interval, and thus, has no
noticeable effect on test particles. The reason for this is that the field
tensor
$f{\alpha \beta} = \partial_{\alpha} a_{\beta} - \partial_{\beta} a_{\alpha}$
contain derivatives of the $\delta$-functions, and when integrated by parts,
a coupling of the $\delta$ to acceleration terms is obtained, causing a large
mass renormalization effect when the test particle hits the surface of
singular support.
This effect actually reduces the impact to a finite value, causing it to
behave as a zero-measure force.

The smooth fields, on the other hand, obey a $1/r^2$ decay power-law. However,
the $(4,1)$ and $(3,2)$ fields differ dramatically in this case.
For the $(4,1)$ metric case, the denominator is positive definite, which
can easily be observed when a non-physical frame of
$n^{\alpha} = [1;0,0,0,0,0]$ is taken
($b^{5} = 0$ in this case, contradicting the theory).
In fact, for any $n^{5} \neq 0$, it causes the field to be a transient
phenomenon, decaying as $1/\tau^2$ for large $\tau \gg \sqrt{\vb{x}^2 + t^2}$.

On the other hand, For the $(3,2)$ case, the fields follow an $O(2,2)$
symmetry as well, which can be seen when $n^{\alpha} = \delta^{\alpha}_{i}$
for one of $i \in \{ 1,2,3 \}$. When this field is integrated over $\tau$,
it produces the $O(2,1)$ GF (proportional to $[t^2 - x^2 - y^2]^{-1/2}$),
which is also the Maxwell field produced by a uniformly moving $3D$
point source in spacelike motion.

In \cite{MCLand1997}, M.C. Land studied the equations of motion of a
test particle in a field with similar singular support behavior. In
particular, the scattering problem in the non-relativistic limit was
derived, in which he noted a failure in matching the well known
Rutherford scattering formula. Land then used the mass-$\tau$
uncertainty relations, similar to the time-energy uncertainty in
non-relativistic QM, to argue that a true point-wise 4D particle is
insufficient to describe a physical source, and thus defined a
\emph{distribution of events} along the $\tau$ parameter, acting
coherently as a single particle. He chose the following distribution
\begin{align}
    j^{\alpha}(x,\tau)
    & =
        \dfrac{b^{\alpha}}{2\lambda}
        \int_{-\infty}^{+\infty} d\tau'
            e^{-|\tau - \tau'| / \lambda}
            \delta^{4}[x - b\tau']
\end{align}
which approaches the point-wise distribution for
$\lambda \rightarrow 0^{+}$, and the Maxwell worldline
(see eq. \eqref{eq:maxwell_current_of_point_particle}) for
$\lambda \rightarrow +\infty$.
Since the fields are linear, the cumulative contribution smoothed out the
$\delta$-function fields. Using numerical computation, Land found
a constraint on $\lambda$.

%% Though the Maxwell current
%% \eqref{eq:maxwell_current_of_point_particle} can be approached by
%% the method of regularization offered by Land, a few other approaches
%% are at hand which may yield similar results. For instance, it is
%% possible \cite{PersonalCommunication} that non-empty space may
%% actually filter very high frequencies observed in the
%% $\delta$-fields. The lowest order zero-mode of the fields is in
%% fact, the Maxwell field (see eq.
%% \eqref{eq:maxwell_field_from_zero_mode}).
We shall show in a subsequent study how this method of regularization
applies to the type of fields we have found here, and make comparison
with observed phenomena.

It was found that the GF's \eqref{eq:green_functions_final} are
consistent with the UMS fields. $\epsilon$ derivative is used to indicate
derivation with respect to the argument, which is maintained even once the
fields are applied on a test particle. Although the derivative seems
to contain a strong distribution $\delta(y)/\sqrt{y}$, this term has
proved essential in the derivation of $\zeta = -1$ smooth fields
\eqref{eq:smooth_fields_through_green_functions}, in which it counter balanced
an infinite contribution from the bounds
$\tau \rightarrow \tau'^{+}_{2}, \tau'^{-}_{1}$. Geometrically, it regularized
the singular support at the 5D light-cone.

The GF's obtained would be used for subsequent studies of radiation-reaction,
2 particle systems and various models of regularizations.
%%
%%
%%
%%
%%
%%
%%
%%
%% END OF CONCLUSIONS
%%
%%
%%
%%
%%
%%

%%%\input{UMP_Bibliography}
%=================================================================%
%-----------------------------------------------------------------%
%- Project...........: UMP - Uniform Motion Paper.                %
%- File..............: UMP_Bibliography.tex                       %
%- Synopsis..........: Bibliography entries for the UMP paper.    %
%-                                                                %
%- Author(s).........: Larry Horwitz and Jigal Aharonovich.       %
%-                                                                %
%-----------------------------------------------------------------%
%=================================================================%

%-----------------------------------------------------------------%
%->>>>>>>>>>>>>>>>>>>>>> END OF DOCUMENT <<<<<<<<<<<<<<<<<<<<<<<<-%
%-----------------------------------------------------------------%
\end{document}